# Non-local dispersion and the reassessment of Richardson's $t^3$-scaling law

G.E. Elsinga[1], T. Ishihara[2] AND J.C.R. Hunt[3]

[1]Laboratory for Aero and Hydrodynamics, Department of Mechanical, Maritime and Materials Engineering, Delft University of Technology, 2628CD Delft, The Netherlands
[2] Graduate School of Environmental and Life Science, Okayama University, Okayama 700-8530, Japan
[3]Department of Earth Sciences, University College London, London WC1E 6BT, United Kingdom

Correspondence to: g.e.elsinga@tudelft.nl

The Richardson scaling law states that the mean square separation of a fluid particle pair grows according to $t^3$ within the inertial range and at intermediate times. The theories predicting this scaling regime assume that the pair separation is within the inertial range and that the dispersion is local, meaning that only eddies at the scale of the separation contribute. These assumptions ignore the structural organization of the turbulent flow into large-scale shear layers, where the intense small-scale motions are bounded by the large-scale energetic motions. Therefore, the large scales contribute to the velocity difference across the small-scale structures. It is shown that, indeed, the pair dispersion inside these layers is highly non-local and approaches Taylor dispersion in a way that is fundamentally different from the Richardson scaling law. Also, the layer's contribution to the overall mean square separation remains significant as the Reynolds number increases. This calls into question the validity of the theoretical assumptions. Moreover, a literature survey reveals that, so far, $t^3$ scaling is not observed for initial separations within the inertial range. We propose that the intermediate pair dispersion regime is a transition region that connects the initial Batchelor- with the final Taylor-dispersion regime. Such a simple interpretation is shown to be consistent with observations, and is able to explain why $t^3$ scaling is found only for one specific initial separation outside the inertial range. Moreover, the model incorporates the observed non-local contribution to the dispersion, because it requires only small-time-scale properties and large-scale properties.

## 1. Introduction

The relative dispersion of two tracer particles in a turbulent flow has received considerable interest, because it is a tractable problem closely connected to turbulent mixing. More specifically, the spatial correlations, including the variance, of a passive scalar are related to the statistics of the relative pair separation (Batchelor 1952, Monin & Yaglom 1975).

If $\boldsymbol{x}_1$ and $\boldsymbol{x}_2$ are the Lagragian trajectories of the two tracer particles, then their relative dispersion is described by the relative separation vector, $\boldsymbol{r}(t) = \boldsymbol{x}_1(t) - \boldsymbol{x}_2(t)$. The separation distance, given by $r(t) = |\boldsymbol{r}(t)|$, is widely believed to scale as $\langle r^2 \rangle \sim t^3$ in the inertial range and for intermediate times, which is referred to as Richardson scaling. Here, $\langle .. \rangle$ indicates averaging over a large ensemble of pairs. Initially, Richardson (1926) obtained the $t^3$ scaling by proposing a diffusion equation for relative dispersion in isotropic turbulence, where, based on his experimental observations, the diffusion coefficient, $K$, was scale dependent according to $K \sim r^{4/3}$.

As Batchelor (1952) noted, the diffusion equation is not exact, but is based on some intuitive or empirical evidence. Moreover, this equation conveniently reduces the full complexity of the turbulent scalar transport to a diffusivity parameter. In Richardson (1926) the diffusivity was a function of the pair separation distance, $r$, while Batchelor (1952) argued it should be a function of time, $t$. Finally, in the Kraichnan (1966) model, the diffusivity depended on both $r$ and $t$. However, all three approaches resulted in a $t^3$ scaling of the mean square pair separation, $\langle r^2 \rangle$, in the inertial range. Nowadays, the diffusion equation is considered unsuitable for modelling the dispersion by a turbulent flow, since turbulence is time correlated and contains non-local effects (e.g. Salazar & Collins 2009, Falkovich, Gawedzki & Vergassola 2001, Thalabard, Krstulovic & Bec 2014). But the $t^3$ scaling law, originally obtained from the diffusion equation, has since been derived by other means as discussed in §2, and is still considered valid.

In their review of pair dispersion, Salazar & Collins (2009) concluded that the Richardson $\langle r^2 \rangle \sim t^3$ scaling law remains unchallenged despite that *"...there has not been an experiment that has unequivocally confirmed Richardson scaling over a broad-enough range of time and with sufficient accuracy."* More than ten years later, this is still the case, as shown in §3. The lack of a clearly observable $t^3$ scaling is commonly attributed to too low Reynolds numbers, too short observation times, mean shear or experimental error, but hardly ever to fundamental limitations in the theory. On the contrary, the ability to predict $t^3$ scaling is often a measure by which a new theory or dispersion model is judged (some examples are given in Sawford (2001)).

The present level of confidence in Richardson scaling is thus largely based on a firm belief in the classical turbulence theory rather than being based on strong empirical evidence. Notwithstanding shortcomings in the simulations and experiments, the lack of convincing evidence calls for a critical reassessment of the underlying assumptions and hypotheses, especially in light of recent advances in the understanding of the structure of the turbulent velocity field.

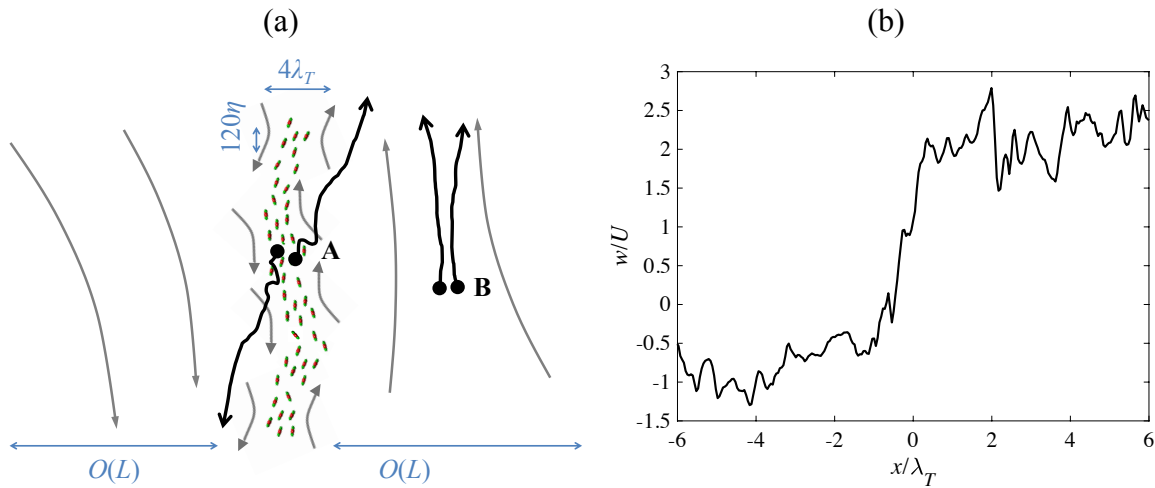

Figure 1. (a) Tracer particle pair dispersion near a significant shear layer, which is characteristic of intermittency at high Reynolds number. Pair A is initially located inside the layer and transported by the small-scale structures within this large-scale layer structure. The small scales are indicated by red and green blobs marking intense dissipation and swirl respectively (see also Elsinga et al. 2017). As soon as the pair leaves the layer, it disperses quickly due to the large velocity difference across the layer. Pair B is located within a large energetic flow region bounding the layer, which is associated with low level velocity fluctuations and slow relative dispersion. (b) Example of a tangential velocity, $w$, profile across a significant shear layer, where $x$ denotes the normal to the layer (data from Ishihara, Kaneda & Hunt (2013)).

Of particular relevance is that high Reynolds number turbulence is highly intermittent, meaning that the most intense small-scale motions are clustered in confined regions of space. Intermittency has been related, at least in part, to the development of shear layer structures, which combine small-scale and large-scale motions (figure 1(a)). Intense small-scale flow structures, i.e. vortices and dissipation sheets, are found within these shear layers, and they scale with the Kolmogorov length scale, $\eta$. The shear layer itself is approximately $4\lambda_T$ thick on average, where $\lambda_T$ is the Taylor length scale, and is bounded on either side by large nearly uniform flow regions that scale with the integral length scale, $L$. The velocity magnitude in the uniform flow regions is of the order of the root-mean-square of a velocity component, $U$. These shear layers are important as they contain significant dissipation (Ishihara, Kaneda & Hunt 2013, Elsinga, Ishihara & Hunt 2020) and affect the average velocity distribution associated with strain (Elsinga et al. 2017). It is this strain that is responsible for the relative dispersion/separation of particle pairs (see Goto & Vassilicos (2004) for an illustration in 2D). Because of their importance and large size, Ishihara, Kaneda & Hunt (2013) have referred to these structures as significant shear layers. Moreover, these layers are persistent and can affect the dispersion over a long time. The lifetime of the significant shear layers is of the same order as that of the large scales, i.e. $L/U$, since the layers are created by large scale regions rubbing against each other. However, the internal structures, such as the small-scale vortices, may have shorter lifetimes.

The observation of significant shear layers challenges the classical understanding of turbulence, which assumes that the small-scales are largely independent of the largest scales. However, these thin shear layers are bounded by large-scale energetic motions, which determine the velocity difference across the layer and the intense small-scale motions within the layer (figure 1(b)). Indeed, Ishihara, Kaneda & Hunt (2013) find that the velocity difference associated with the intense small-scale structures is of the order of $U$. Furthermore, significant energy transfer was found between the largest and smallest scales, and vice versa, near the significant shear layer (Aoyama et al. 2005), which suggests that the scales are strongly mutually dependent. Since these shear layers contribute significantly to the turbulent strain and dissipation, as mentioned above, they are likely to affect pair dispersion in non-classical ways. This is confirmed by new results for pair dispersion inside a significant shear layer, which are presented in §3.3.

The aim of the present paper is to open a discussion on the validity of the $\langle r^2 \rangle \sim t^3$ scaling law and its underlying assumptions in particular (§2). Furthermore, numerical and experimental results reported in the literature since the 2009 review by Salazar & Collins are examined for possible evidence of $t^3$ scaling in the ranges predicted by the theory (§3). An alternative model for pair dispersion at intermediate times is introduced in §4. This model captures features of the mean square separation currently unexplained by theory. The main findings and observations are summarized in §5.

As a final introductory remark, we briefly comment on the other scaling regimes in pair dispersion. The initial ballistic (or Batchelor) regime is kinematic in nature and assumes that the particles maintain their initial velocity over short time-scales. This approximation is valid for any continuous velocity field in the limit of small $t$, and results in a $\langle r^2 \rangle \sim t^2$ scaling law (Batchelor 1950). For long times, the pair separation distance increases beyond the integral length scale, where the particles move independently leading to a diffusive dispersion regime, which is known as the Taylor regime (Taylor 1922). Both the initial short-time Batchelor and the long-time diffusive regime are well established and are independent of the detailed structure of turbulence, which is not the case for the intermediate-time Richardson scaling regime considered here. We return to these other regimes in §4.

## 2. Derivations of $t^3$ scaling

This section reviews a number of different theories, which predict a $\langle r^2 \rangle \sim t^3$ scaling regime (§§2.1-2.3). Particular emphasis will be on the ranges for which $t^3$ scaling is predicted. It is intended as an overview of the broad variety of approaches without claiming to be complete. However, in the end, all these theoretical studies use quite similar assumptions. Their validity is discussed in §§2.4 and 2.5.

### *2.1 Dimensional analysis*

Batchelor (1950) obtained $t^3$ scaling for initial separations, $r_0 = r(t=0)$, in the inertial range from dimensional arguments, which are similar to those originally presented by Obukhov (1941). Within the inertial range the relative motion (i.e. velocity) is said not to be affected by viscosity. Furthermore, the dependence of the dispersion on the initial separation, $r_0$, and the energy containing motions is assumed negligible at intermediate times (defined in §3.1) when $r_0^2 \ll \langle r(t)^2 \rangle \ll L^2$. In the spirit of K41 (Kolmogorov 1941), the relative motion then depends only on the mean dissipation rate, $\varepsilon$, and time, $t$. From dimensional consistency, it follows that (Batchelor 1950):

$$\frac{d\langle r(t)^2 \rangle}{dt} = C\varepsilon t^2 \qquad (2.1)$$

where $C$ is a dimensionless constant. Integration subsequently yields:

$$\langle r(t)^2 \rangle - r_0^2 = g\varepsilon t^3 \qquad (2.2)$$

Here $g$ ($= C/3$) is the Richardson constant. In a later paper, Batchelor (1952) introduced a time shift $t_1$ allowing for some influence of the initial pair separation $r_0$ on the effective origin ($t_1$, $r_1^2$), which resulted in:

$$\langle r(t)^2 \rangle - r_1^2 = g\varepsilon(t - t_1)^3 \qquad (2.3)$$

Other than that, the set of assumptions resulting in (2.3) remained the same as for (2.2). Further justification for an effective origin is given by Ishihara & Kaneda (2002), which is based on the observation of a linear region in the Lagrangian correlation of the velocity difference.

### *2.2 Mechanistic/stochastic approach*

A physical mechanism leading to $t^3$ scaling was proposed by Bourgoin (2015). In this model, the mean square particle separation, $\langle r(t)^2 \rangle$, is described by a stepwise ballistic growth. During each step $k$, the pair's relative velocity squared is constant and given by the second-order Eulerian structure function, $S_2$, evaluated at the scale of the particle separation distance, $r_k$. This leads to ballistic dispersion, which is maintained for a time period $t_k$. Accordingly, the pair separation will have grown to $r_{k+1}$ at the start of the next step $k+1$. The ballistic process repeats at the new scale, $r_{k+1}$, with the associated $S_2(r_{k+1})$ and time scale $t_{k+1}$. Both the Eulerian structure function and the time scale are related to the particle separation distance by imposing K41 scaling. Specifically, $S_2 \sim (\varepsilon r)^{2/3}$, which applies in the inertial range. This implies that the theory is valid for $r_k$ (including $r_0$) within the inertial range. Furthermore, $t_k = \alpha S_2(r_k)/2\varepsilon$, where $\alpha$ is referred to as a persistence parameter. The condition $\alpha = 1$ corresponds to the eddy turnover time at scale $r_k$. This stepwise process leads to:

$$\langle r(t)^2 \rangle = g\varepsilon(t - t_0)^3 \qquad (2.4)$$

where the virtual time origin $t_0$ depends on the initial separation. The model parameter $\alpha$ can be tuned to yield a Richardson constant $g = 0.55$ consistent with reported values in the literature.

It is important to note that the ballistic dispersion model is local in its original implementation, that is, only the eddies of scale $r_k$ contribute to the relative velocity and these eddies have inertial range scaling properties. Furthermore, the variations in the square separation at $t > 0$ are not considered, i.e. the model only uses the mean.

The ballistic approach is general and can be adapted to turbulent shear flow (Polanco et al. 2018). Moreover, the actual $S_2$ (as opposed the inertial range model) may be inserted, which yields deviations from equation (2.4) (Liot et al. 2019, see also §4). However, we focus on the case of homogeneous isotropic turbulence with inertial range and local dispersion assumptions, because it yields the classical Richardson scaling, which is of interest here.

Another ballistic model was proposed by Thalabard, Krstulovic & Bec (2014), who considered the pair dispersion as a continuous-time random walk process. The inertial range and local dispersion assumptions are the same as in Bourgoin (2015), however, the stepwise ballistic scenario is formulated in terms of increments in the pair separation distance, $r$, (as opposed to its square) and a different choice is made for the relative velocity between the particles at a given scale. These choices also lead to a $t^3$ scaling for $\langle r(t)^2 \rangle$ in the inertial range, but the separation growth is governed at leading order by third-order velocity increments as compared to the second-order increments in the Bourgoin (2015) model. Thus, the resulting physics is different due to different choices for the relative velocity between the particles. Related continuous-time random walk descriptions of pair dispersion include the so-called Lévy walks (e.g. Shlesinger, West & Klafter 1987), in which a probability distribution is assumed for the steps in the pair separation distance and the associated time steps.

In Lagrangian stochastic models (Thomson 1987, Wilson & Sawford 1996, Sawford 2001), the dispersion of particles is treated as a Markov process, where the changes in a pair's relative separation and relative velocity depend only the present separation and relative velocity. Therefore, the process is memoryless and ignores any history effects. Then, the relative velocity increment at each time step is described by a stochastic differential equation containing a drift term and a diffusion term. The latter adds a Gaussian white noise, which amplitude scales with $(\varepsilon \, dt)^{1/2}$ in the inertial range according to K41. However, a dissipation range correction has been considered by Borgas & Yeung (2004). The drift term can be related to the Eulerian relative velocity probability-density-function, $p_E$ (Thomson 1987). Assuming K41 inertial range properties for $p_E$, that is, the shape of $p_E$ is self-similar and its variance scales according to $\sim(\varepsilon r)^{2/3}$, results in Richardson scaling, see Sawford (2001) for a review and Devenish & Thomson (2019) for a recent example. Note, however, the large scatter in the Richardson constants predicted by the different Lagrangian stochastic models (Sawford 2001). Clearly other modelling aspects also play a critical role. The above assumes that $r$ (while distributed) remains within the inertial range, which is a local dispersion assumption in the sense that the small dissipative scales and the large scales do not influence the dispersion. In an actual turbulent flow, $p_E$ is not self-similar (Sawford & Yeung 2010), which leads to deviations from the classical Richardson scaling as discussed further in §4.

### *2.3 Spectral approach*

Malik (2018) related the slope of the kinetic energy spectrum, $E(k)$, to the pair dispersion power law. Note that there was no dissipative range in the assumed energy spectrum. Consequently, the initial separation was always in the inertial range. For local dispersion, i.e. dispersion by the turbulent scales equal to the pair separation, and a $k^{-5/3}$ energy spectrum, the Richardson $t^3$ scaling law was obtained. However, including effects of non-local scales, that is, inertial range scales larger than the separation distance, was shown to enhance the dispersion and increase the power scaling ($t^\gamma$ with $\gamma > 3$). So, the assumption of local dispersion seems important in obtaining $t^3$ scaling.

*2.4 Why Kolmogorov theory works for average energy related statistics only*

The existing approaches predicting $t^3$ scaling (§§2.1-2.3) rely on the existence of an inertial range, where all flow properties depend only on $\varepsilon$ and the *local* scale (length scale $r$ or time scale $t$). This is referred to as Kolmogorov theory, or K41, and represents the classic inertial range assumption. K41 predicts, among other things, that, in the inertial range, $S_2 \sim (\varepsilon r)^{2/3}$ and $E \sim k^{-5/3}$. These results were used in §2.2 and §2.3 respectively. While $S_2 \sim (\varepsilon r)^{2/3}$ and $E \sim k^{-5/3}$ appear to be accurate subject to minor corrections (e.g. Donzis & Sreenivasan 2010), this does not imply that K41 can be extended to other turbulence properties, including dispersion, without question. We explain our reservations below.

In order to better understand the successes and limitations of K41, it is useful to consider the intermediate range as a matching or overlap region, which connects the small $r$ regime to the large $r$ regime. The present matching argument developed for structure functions is similar to that of Tennekes & Lumley (1972, p. 264) and George (2013) for the inertial range in the energy spectrum. Consider the $n^{th}$ order velocity structure function, $S_n(r) = \langle |\delta \boldsymbol{u}(r,0)|^n \rangle$, where $\delta \boldsymbol{u}$ is the relative velocity between the particles at $t = 0$. Assume that for small separation $r$, these structure functions scale with the Kolmogorov length and velocity scales, $\eta$ and $u_\eta$ respectively, which is written as:

$$S_n(r) = u_\eta^n S_n^+(r^+) \tag{2.5}$$

where $r^+ = r/\eta$. At large $r$, integral scaling applies, which is written as:

$$S_n(r) = U^n \widetilde{S_n}(\tilde{r}) \tag{2.6}$$

where $\tilde{r} = r/L$. Furthermore, assume that there exists an overlap or matching region where both scalings are valid. In that case, the derivatives $dS_n/dr$ from equations (2.5) and (2.6) can be equated, which results in:

$$(r^+)^{1-n/3} \frac{dS_n^+}{dr^+} = D^{-n/3} (\tilde{r})^{1-n/3} \frac{d\widetilde{S_n}}{d\tilde{r}} \tag{2.7}$$

In the above equation, both sides have been multiplied by $r^{1-n/3}$ and the relation $u_\eta^3/\eta \equiv \varepsilon = DU^3/L$ has been used, where the normalized dissipation rate, $D$, is a constant in (near) equilibrium turbulence at sufficiently large Reynolds number ($Re_\lambda > 100$, Sreenivasan 1998, Kaneda et al. 2003). The latter follows from the energy balance, where the average dissipation-rate must be equal to the average turbulent kinetic energy production-rate when turbulence is at equilibrium. Note that the energy balance $\varepsilon = DU^3/L$ does not require an inertial range (see also Pope 2000). It only assumes that production occurs at large scales, and hence the production-rate scales with $U^3/L$. In the limit of large Reynolds number, and hence $r^+ \to \infty$ and $\tilde{r} \to 0$, the left and right-hand-side of equation (2.7) must be equal to the same constant, which we define as $\frac{n}{3} C_n$. Subsequent integration yields:

$$S_n(r) = C_n (\varepsilon r)^{n/3} \tag{2.8}$$

for the overlap region. Equation (2.8) presents the classical Kolmogorov relations. So, if the assumption presented in equation (2.5) holds, we expect (2.8), and hence K41, to be valid for the intermediate range in turbulence at sufficiently large Reynolds number. The validity of (2.5) is easily verified.

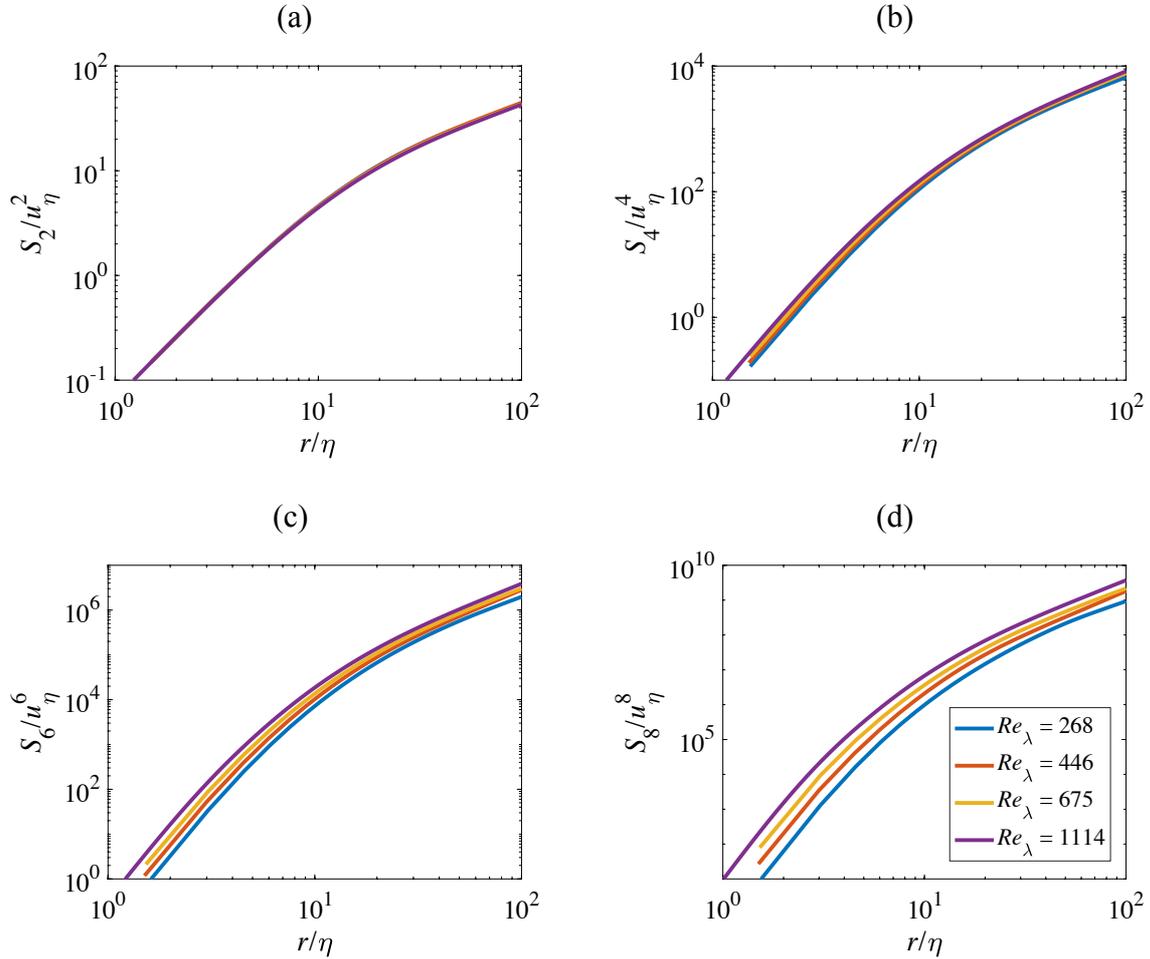

Figure 2. Longitudinal velocity structure functions of order 2, 4, 6 and 8 ((a)-(d) respectively) at $r/\eta < 100$, which is the length scale for small-scale coherence (§3.1). The structure functions are normalized using the Kolmogorov velocity and length scales. DNS results are presented for four different Reynolds numbers (see legend inset in (d)). The DNS data sources are Ishihara et al. (2007) ($Re_\lambda$ = 268, 446 and 675, cases 1024-2, 2048-2 and 4096-2 in their paper) and Elsinga, Ishihara & Hunt (2020) ($Re_\lambda$ = 1114, case 8192-2 in their paper, however using a slightly different time instant).

For the special case of the second-order structure function ($n$ = 2), Kolmogorov scaling applies to a very good approximation at small $r$ (figure 2(a)). Consequently, $\varepsilon$ and $r$ appear as the only relevant parameters in the overlap region of $S_2$ consistent with Kolmogorov theory (equation (2.8)). However, the collapse of $S_2$ with $u_\eta$ and $\eta$ is to be expected owing to the $\varepsilon$-based definitions of the Kolmogorov scales. Since $\varepsilon \sim \nu \langle (du/dr)^2 \rangle = \lim_{r \downarrow 0}(\nu S_2/r^2)$ and $\varepsilon = \nu(u_\eta/\eta)^2$, it follows that $S_2 \sim (u_\eta r/\eta)^2$ at small $r$, which validates equation (2.5) for $n$ = 2.

Generally, Kolmogorov scaling does not collapse velocity structure functions at small $r$. Reynolds number dependencies are clearly visible for $n \geq 6$ (figures 2(c-d)), and may already be noticed at $n$ = 4 (figure 2(b)). This means that another, independent scale needs to be combined with the Kolmogorov scales in order to collapse the data. The only remaining independent turbulent scale (in the current understanding of turbulence) is the integral scale. Therefore, the lack of a collapse in figure 2 suggests that large scale influences are present at small $r$. Indeed, the large scales influence the magnitude and the size of the most intense dissipation and velocity gradients (Elsinga, Ishihara & Hunt 2020). Also, Yeung, Brasseur, & Wang (1995) found evidence of a strong and direct coupling between the large and the small scales of motion in both physical and Fourier space, which resulted in a quick response of the small-scales to changes at the largest scale. These couplings may be associated with large energy-containing motions bounding the small-scale motions within the significant shear layer structures (Ishihara, Kaneda & Hunt 2013, Elsinga et al. 2017). Moreover, a dependence on the larger scales may be inferred from the anisotropy of the small scales (Shen & Warhaft 2000, La Porta et al. 2001, Fiscaletti et al. 2016). The above observations are clearly inconsistent with the classical assumption that the small scales are independent of the large scales apart from the mean energy transferred (Kolmogorov 1941). We should mention that these large-scale influences at small $r$ do not appear in $S_2$ (figure 2(a)), because of the way $u_\eta$ and $\eta$ have been defined (see above). The lack of collapse at small $r$ in figures 2(c-d) causes equation (2.8) to fail in predicting the power law exponent at intermediate $r$ when $n > 5$ (Benzi et al. 1993, their figure 2). Large-scale influences, i.e. a contribution from $U$ at small length scales, can explain the observed discrepancy in these power law exponents (She & Leveque 1994).

In summary, Kolmogorov theory appears to be suitable for statistical averages closely associated with energy (e.g. energy spectrum, $S_2$, $D$ and its Lagrangian equivalent $D_L$), because K41 is based on the energy balance (and equilibrium). However, Kolmogorov theory is less suitable for flow properties other than energy, such as the higher order moments of velocity difference ($n > 5$ in particular), because, in those cases, the large-scale influences at small $r$ are not accurately accounted for. Note that the intermediate region may still be considered as an overlap region, but velocity and length scales different from $u_\eta$ and $\eta$ are required to collapse $S_n$ at small $r$ and $n \gtrsim 5$. Different scales may also improve the collapse for lower order moments, $n \sim 4$, but the gain will be less obvious, since the observed deviations from K41 are quite small in those cases.

What are the implications for pair dispersion? For the initial Batchelor regime at small $t$, the mean square separation depends on $S_2(r_0)$ (Batchelor 1950, §4), and hence collapses in Kolmogorov scaling for a given $r_0$ (e.g. Sawford, Yeung & Hackl 2008, Salazar & Collins 2009). So, the small $t$ regime depends on $u_\eta$, $\eta$ and $r_0$. The additional $r_0$ dependence may carry over into the overlap region and affect the scaling exponent at intermediate time (and indeed it does, §§3 and 4). This is similar to what has been observed for the velocity structure functions, where the Kolmogorov scales fail to collapse the higher order structure functions at small $r$ causing the breakdown of K41 in the intermediate range for $n \gtrsim 5$. The additional $r_0$ dependence may also introduce a Reynolds number dependence in the scaling exponent. Such an $r_0$ dependence at intermediate times can be viewed as a non-local effect or a history effect, which is not included in the K41-based theories predicting Richardson scaling (§§2.1-2.3). Moreover, $r(t)$ is distributed at the end of the Batchelor regime, and hence at the start of the intermediate regime, which further complicates the analysis as discussed in §2.5. It is therefore not obvious that K41 should apply to pair dispersion at intermediate times.

*2.5 Non-local dispersion and turbulent structure*

The approaches in §§2.2-2.3 include the idea that, at any given time, only the turbulent eddies at the scale of the mean pair separation distance, $\langle r(t)^2 \rangle^{1/2}$, contribute to the rate of change of the separation distance. This is known as local dispersion, as opposed to non-local dispersion, where the rate of change of the pair separation distance is affected by turbulent eddies at scales other than $\langle r(t)^2 \rangle^{1/2}$.

Dispersion, however, contains non-local contributions. For example, some particles, initially at $r_0$ within the inertial range, are brought closer together as time progresses with $r$ eventually entering into the small-scale range, while other pairs move far apart causing $r$ to be in the large-scale range. Consequently, the pair separation probability-density-function (PDF) is very broad (e.g. Scatamacchia, Biferale & Toschi 2012, Bitane, Homann & Bec 2013), which means that the pair dispersion for $r_0$ at later time is affected by energetic scales and viscous scales simultaneously. Furthermore, the PDFs may depend on $r_0$, which could help to explain a $r_0$ dependence at intermediate time. These issues are minor when the Reynolds number is extremely large and it would take any pair very long to disperse to scales well outside the inertial range. However, as argued in §2.4, notable large-scale influences persist down to small $r$. This makes predicting the tails of the PDF, and hence the mean square separation, highly complex.

The complex relation between dispersion and turbulent scales is illustrated in figure 1. A particle pair leaving a small-scale eddy inside the significant shear layer almost immediately enters the neighboring energetic scales (pair A in figure 1(a)). In that case, the dispersion is always affected by the viscosity or the energy-containing eddies, and there is no important contribution from inertial range eddies, even when $r$ is in the inertial range. Furthermore, the energetic large-scale motions bounding the significant shear layer determine the velocity difference across the layer and thereby they control the magnitude of the small-scale eddies and the small-scale dispersion inside the layer. These are significant non-local effects, which are assessed in §3.3.

Turbulent structure also contributes to the selective sampling of the flow. Pairs that are randomly distributed initially (isotropy) align and cluster due to flow structure. For example, pairs cluster onto sheets around a shear layer flow structure or a node-saddle topology (e.g. Goudar & Elsinga 2018). All pairs then disperse along the directions of the sheet, which roughly correspond to the directions of extensive strain. The approximate alignment between the most extensive strain and the direction of large elongation between particle pairs has also been noted by Devenish & Thomson (2013). Consequently, the particles probe the turbulent flow along those specific directions, which may have a different velocity distribution across the scales as compared to the unconditional/isotropic energy distribution. For instance, the kinetic energy spectrum along the most extensive straining direction of an average shear layer structure is different ($k^{-1}$) from the usual $k^{-5/3}$, which is the average over all directions (Elsinga & Marusic 2016). These effects are not well understood at present, but may have important implications. A related discussion for single particle statistics can be found in Lalescu & Wilczek (2018).

## 3. Observations

Now that the theory has been reviewed and the issues concerning the assumptions have been discussed, we turn our attention to the evidence for Richardson scaling. Results from DNS (§3.2), kinematic simulations (§3.4) and experiments (§3.5) are considered. Furthermore, the contribution from the significant shear layers to the overall pair dispersion statistics is examined in §3.3. However, first the requirements are given for positively identifying a Richardson scaling regime (§3.1).

### 3.1 Requirements for testing Richardson scaling

When testing for Richardson scaling, the initial separation $r_0$ is required to be in the inertial range consistent with the assumptions made when deriving the scaling law (§2). Typically, this requirement is translated into $\eta \ll r_0 \ll L$ (e.g. Sawford (2001) and Salazar & Collins (2009)), which is correct but imprecise. Often it is misinterpreted as $r_0 \sim 10\eta$ being sufficient, simply because it is an order of magnitude larger than the lower bound, $\eta$ (see §§3.2 and 3.5). Note that $r_0 \sim 10\eta$ is still within the linear core of the vortices and the dissipation sheets (Jiménez et al. 1993, Elsinga et al. 2017), which are small-scale structures. The actual lower bound, however, is provided by the assumption that the relative velocity at scale $r_0$ is not affected by viscosity (see §2), i.e. not affected by the small dissipative scales. Though it is difficult to define the lower bound for the inertial range exactly, we may specify that $r_0$ should be larger than the characteristic size of the dissipation structures, which has been determined at $\sim 60\eta$ (Elsinga et al. 2017). Moreover, the dissipation spectrum drops quickly for dimensionless wavenumbers $k\eta < 10^{-1}$ (e.g. Pope 2000) corresponding to a physical length scales larger than approximately $60\eta$. A more conservative criterion would be $r_0 \gtrsim 120\eta$, which is based on the coherence length of small-scale vorticity (Elsinga et al. 2017). This stricter criterion is consistent with the second-order Eulerian structure function revealing a $r^{2/3}$ inertial range scaling for separation distances larger than $100\eta$ (e.g. Donzis & Sreenivasan 2010). At present, there is limited data available for $r_0 > 60\eta$, let alone $r_0 > 120\eta$, making it difficult to effectively assess the scaling law for the inertial range. The exception may be the experiments discussed in §3.5. Furthermore, $t^3$ scaling should appear for a wide range of initial separations, $r_0$, within the inertial range.

Concerning the temporal range, the $t^3$ scaling is predicted for intermediate times, after which the initial condition is forgotten (§2.1) and $r_0^2 \ll \langle r(t)^2 \rangle \ll L^2$. This intermediate range is expected to lie somewhere between the Batchelor time scale, $t_B = r_0^{2/3} \varepsilon^{-1/3}$, and the Lagrangian integral time scale $T_L$ (Batchelor 1950). Furthermore, the $t^3$ scaling should appear for a decade of $t$ in order to be convincing, as explained below.

Alternatively, it has been suggested that Richardson scaling can also appear for very small initial separations, $r_0 \lesssim \eta$, and intermediate times, $t \gg \tau_\eta$, when $\eta \ll r \ll L$ (Batchelor 1952, Monin & Yaglom 1975). In that case, the effect of viscosity is lost and it is hypothesized that the particle pairs ultimately forget their $r_0$ before entering the Taylor regime (Batchelor 1952). In this scenario, the bulk of the pairs need to reach the inertial range ($r \gtrsim 120\eta$) first. During this time, the initial condition will not be forgotten due to the spatial coherence that exists up to $120\eta$ (Elsinga et al. 2017). Because the velocity of the tracers and the fluid is the same, it is reasonable to assume that the pair travel time and the turn-over time of the fluid structure (and hence its decorrelation time) are similar for a given length. In that case, spatial coherence of a flow structure implies sufficient temporal coherence. Once the pairs enter the inertial range, the $r_0$-dependence may be gradually lost in a similar process as for the pairs with $r_0$ inside the inertial range. Because of the additional time needed to reach the inertial range when $r_0 \lesssim \eta$, we expect that Richardson scaling appears first for $\eta \ll r_0 \ll L$, which, moreover, is the common condition appearing in reviews on the subject (Sawford 2001, Salazar & Collins 2009). In any case, if a true Richardson scaling regime exists for $r_0 \lesssim \eta$, then there is no reason why it should not appear for $\eta \ll r_0 \ll L$ from the theoretical point of view (both conditions neglect the effects of viscosity and $r_0$ after some time when $r$ is within the inertial range). So far, ballistic and Lagrangian stochastic modelling frameworks (§2.2) have not been able to confirm exact $t^3$ scaling when $r_0$ is within the dissipative range. This is discussed in more detail in the last paragraph of §4. Also, the spectral approach (§2.3) has not yet included a dissipative range. Therefore, we focus on the condition for $\eta \ll r_0 \ll L$, which has received wider theoretical support until now (§2) and may reveal Richardson scaling first. However, results for $r_0 \lesssim \eta$ are commented on when relevant.

Finally, there is the issue of the functional form of the Richardson scaling regime (equations (2.1)-(2.4) or the one that is often used in practice: $\langle |\boldsymbol{r}(t) - \boldsymbol{r}(0)|^2 \rangle = g\varepsilon t^3$). For large times, i.e. $t \gg t_0$ and $\langle r(t)^2 \rangle \gg r_0^2$, these relations are equivalent. However, the observation time in the presently available simulations and experiments is limited due to the limited Reynolds numbers achieved. Therefore, we briefly comment on the issue. The most general form is given by equation (2.3) and uses an effective, or virtual, origin ($t_1, r_1^2$). The other forms can be obtained by introducing specific choices for $t_1$ and $r_1^2$. The use of a virtual origin has received important criticism, because it allows fitting any power law to the data (e.g. Ouellette 2006, Salazar & Collins 2009). This is illustrated in figure 3. Here, the actual mean square separation is taken to evolve according to $\langle r(t)^2 \rangle / \eta^2 = (t/\tau_\eta)^2$ (blue line), where $\tau_\eta$ is the Kolmogorov time scale. However, by introducing a virtual origin when plotting the exact same data, we can observe approximate $t^3$ scaling over a short temporal range, even if the actual dependence is $t^2$. The plot also shows that for sufficiently long times, the true $t^2$ scaling is recovered for all cases (($t-t_1$)>100$\tau_\eta$ in the example of figure 3). In this way, a spurious $t^3$ scaling range can be obtained for at most half a decade (approximately ¼ decade due to a time shift and another ¼ decade due to an $r_1^2$ shift). Hence, the issue of the virtual origin appears irrelevant if a $t^3$ scaling range can be observed for at least a full decade of time. In that case, we may admit the use of the most general form with the virtual origin as a fit parameter (equation (2.3)). The requirements for identifying Richardson scaling in a linear plot of $\langle r(t)^2 \rangle^{1/3}$ versus time are similar, as discussed in Appendix A.

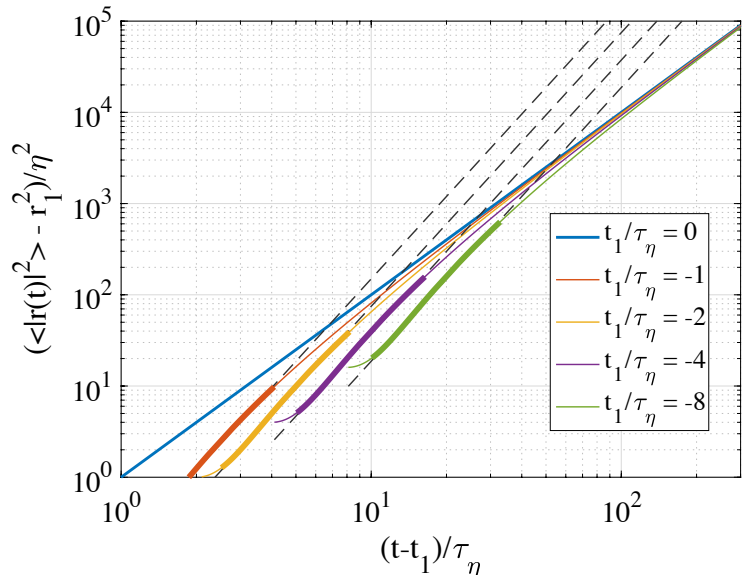

Figure 3. The mean square separation versus time, where the input is an artificial dispersion that evolves according to $\langle r(t)^2 \rangle / \eta^2 = (t/\tau_\eta)^2$ (blue line, $t_1$=0, $r_1$=0). The other lines show the exact same data, but plotted applying different virtual origins ($t_1, r_1^2$). This leads to a spurious $t^3$ scaling range over half a decade of time (marked by the thick lines). Dashed lines indicate $t^3$ power laws for reference. Note that for all cases the correct $t^2$ scaling is recovered at large times.

### 3.2 Results from numerical simulations

At the time of the review of Salazar & Collins (2009), the DNS of particle pair dispersion in homogenous isotropic turbulence (HIT) was available for $Re_\lambda$ up to 650 (Sawford, Yeung & Hackl 2008). These simulations have not confirmed Richardson scaling beyond any doubt, as also mentioned in their review. Either approximate $t^3$ scaling was obtained for a single initial separation outside the inertial range ($r_0 < 60\eta$), or observed only for very short times, much less than a decade. However, the Reynolds number in numerical simulations has increased thereby expanding the inertial range that can be probed.

Bitane, Homann & Bec (2013) considered HIT at $Re_\lambda$ = 460 and 730, and presented mean square separation, $\langle r(t)^2 \rangle$, data for $r_0 \leq 24\eta$ (their figure 2). For $r_0 \approx 4\eta$ and $t > 10\tau_\eta$, a $t^3$ regime is observed for almost a decade before the separation reaches the integral scale, while the exponent is larger and smaller for $r_0 < 4\eta$ and $r_0 > 4\eta$ respectively. Similar results were obtained by Bragg, Ireland & Collins (2016) at $Re_\lambda$ = 582. Assuming these results can be extrapolated to the inertial range (i.e. $r_0 > 60\eta$), the exponent is expected to further decrease, away from the predicted value of 3.

In a related paper (Bitane, Homann & Bec 2012), a short time $t^2$ correction term to the Richardson regime was proposed, which depended on $r_0$. However, the data (especially for large $r_0$) did not extend up to times where the correction is negligible in order to have a true $t^3$ scaling. Furthermore, the correction term was found to be zero only for $r_0 \approx 4\eta$ consistent with their 2013 results discussed above. Hence, they refer to $r_0 \approx 4\eta$ as the 'optimal choice' for observing $t^3$ behavior. As remarked before (§3.1), at this small initial separation, the pair is released within the same small-scale flow structure and not within the inertial range. Moreover, true Richardson dispersion applies to a range of initial separations, as opposed to one specific value of $r_0$.

Additionally, Bitane, Homann & Bec (2012, 2013) define a new time scale, $t_t = S_2(r_0)/(2\varepsilon)$, for the end of the initial ballistic $t^2$ regime, after which a Richardson regime could appear. Here, $S_2(r) = \langle |\delta \boldsymbol{u}|^2 \rangle$ is the second-order Eulerian structure function, and $\delta \boldsymbol{u}$ is the longitudinal velocity difference over a distance $r$. The time scale $t_t$ was shown to collapse the end of the Batchelor regime for their data at $Re_\lambda$ = 460 and 730 (Bitane, Homann & Bec 2013).

An even higher Reynolds number, $Re_\lambda$ = 1000, was achieved by Buaria, Sawford & Yeung (2015). Moreover, they presented mean square separation data for the inertial range, that is $r_0 = 64\eta$ and $256\eta$, which are reproduced in figure 4 (blue solid lines) along with the results for the other initial separations (red dashed lines). Note that the $t^2$ compensated mean square separation is shown here, such that horizontal lines indicate Batchelor $t^2$ scaling. A Richardson $t^3$ power law is indicated by a dash-dotted line for reference. Again, a $t^3$ regime is observed only for $r_0 \approx 4\eta$, while the larger (up to $256\eta$) and the smaller initial separation results approach this line from above and below respectively. However, the results for these other $r_0$ do not reveal an exact $t^3$ scaling. Focussing on the inertial range (blue curves), it is seen that after the initial Batchelor $t^2$ scaling regime, the compensated mean square separation first deceases (implying $t^\beta$ scaling with $\beta < 2$) and reaches a minimum. The time scale associated with this decrease appears to be the Batchelor time scale (marked + in the plot). Moreover, the mean square separation at this point is of the order of $4\lambda_T$ (marked by dotted line with circles in the plot), meaning that a significant number of pairs will have separated by more than the thickness of the significant shear layer. These pairs may already feel the effect of the energetic large-scale motions (figure 1). Following the minimum, at around $t = t_t$ (* in figure 4), the compensated mean square separation increases, but does not reach a $t^3$ regime. Furthermore, the mean square separation quickly reaches integral length scales. For $r_0$ within the inertial range, there is half a decade, or less, between $t_t$ and the time corresponding to the condition $\langle |\boldsymbol{r}(t) - \boldsymbol{r}(0)|^2 \rangle = L^2$ (indicated by dotted line with squares). When the mean square separation increases beyond $L^2$, the curves for all initial separations seem to converge to a common point (figure 4). From that common point the onset of a Taylor regime is anticipated, where the influence of the initial separation is lost. The data show that the mean square separation remains dependent on $r_0$ up to the simulated time, while the extrapolated lines suggest that this dependence persists until the Taylor regime is approached. The $r_0$–dependence is a clear violation of the assumptions leading to a Richardson scaling regime for $r_0$ within the inertial range, as well as for $r_0 \lesssim \eta$ (§3.1). In conclusion, at $Re_\lambda$ = 1000 there is still no sign of a true Richardson scaling regime for the inertial range, and large-scale influences appear already at the end of the Batchelor regime when the pair separation is of the order of the significant shear layer thickness.

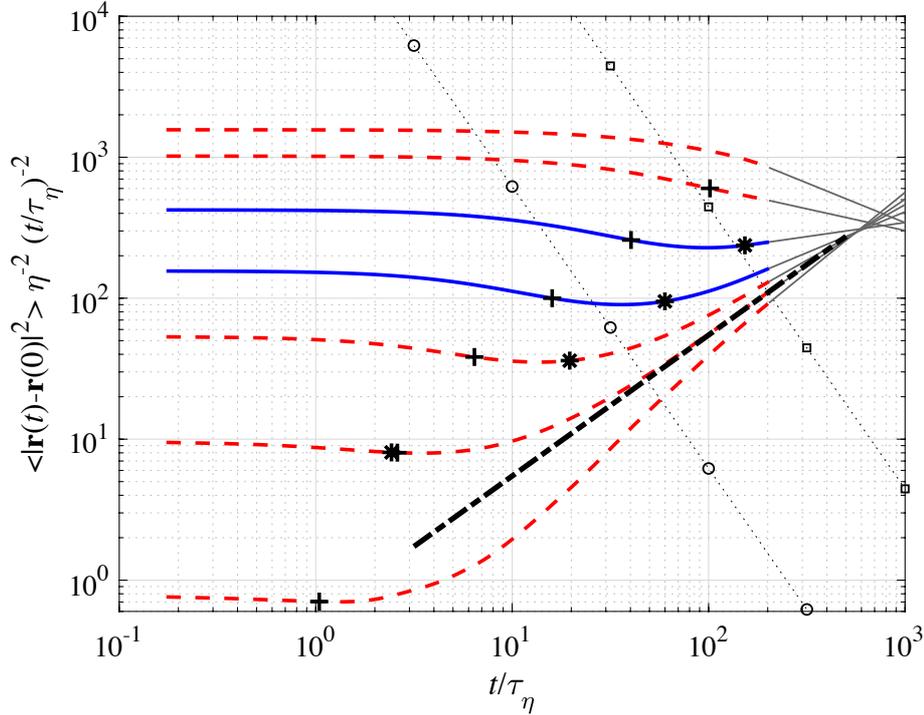

Figure 4. The mean square separation versus time at $Re_\lambda = 1000$, data from Buaria, Sawford & Yeung (2015), where $T_L/\tau_\eta \approx 80$. The thick dashed and solid lines are for different initial separations, $r_0/\eta = 1, 4, 16, 64, 256, 1024$ and $4096$ (increasing upwards), where the inertial range ($r_0/\eta = 64$ and $256$) is marked by the blue solid lines. Note that the mean square separation is multiplied by $t^{-2}$, such that a horizontal line corresponds to Batchelor scaling, while the slope representing the Richardson $t^3$ scaling is indicated by the dash-dotted line. The relevant time scales for each initial separation are marked by symbols, where (+) indicates the Batchelor time scale $t_B$, and (*) indicates $t_t$. Important length scales are indicated by black dotted lines with symbols, where (circles) mark the condition $\langle |r(t) - r(0)|^2 \rangle = (4\lambda_T)^2$ and (squares) mark $\langle |r(t) - r(0)|^2 \rangle = L^2$. The thin grey lines for $t/\tau_\eta > 200$ are extrapolations from the data suggesting a convergence towards a common point. Beyond this point, the Taylor dispersion regime is anticipated (§4).

At $Re_\lambda = 1000$, inertial range scaling is observed in the energy spectrum and the second-order velocity structure function to a good approximation over 1–1.5 decades (e.g. Donzis & Sreenivasan 2010, Ishihara et al. 2016, 2020). This is consistent with defining an approximate inertial range as $60\eta < r < L$ ($\approx 2100\eta$ at $Re_\lambda = 1000$), see (§3.1). However, inertial range scaling in these statistics does not imply inertial range scaling in pair dispersion, i.e. Richardson scaling, as argued in §2.4.

As mentioned, the DNS data reveal a Reynolds number dependence of the mean square separation at intermediate times, which is most notable for $r_0 \gtrsim 16\eta$. Figure 5 presents results from Buaria, Sawford & Yeung (2015) for $r_0 = 16\eta$ and $64\eta$ as an example. Furthermore, the result for $r_0 = 4\eta$ is included as a reference. The initial deviation from the Batchelor regime seems to collapse in Kolmogorov scaling. However, in the intermediate range beyond $t/\tau_\eta \approx 60$, the slope, and hence the scaling exponent, is found to change with the Reynolds number for $r_0 = 16\eta$ and $64\eta$ (figure 5(b) and (c)). Note that the intermediate range is relatively short at these Reynolds numbers. If we strictly define the intermediate range for $r_0 = 4\eta$ as the range where the $t^3$ compensated mean square separation is nearly constant between the values 0.55 and 0.57 (figure 5(a)), then the intermediate range for $r_0 = 4\eta$ starts at $t/\tau_\eta \approx 60$ for all Reynolds numbers and extends to $t/\tau_\eta \approx 80$, 100 and 180 for $Re_\lambda = 390$, 650 and 1000 respectively. Here, it has been assumed that $t^3$ scaling is attained for $r_0 = 4\eta$, which is in line with observations as discussed above. Clearly, the intermediate ranges are very short. The intermediate ranges for the other $r_0$ are more difficult to infer, because their scaling exponents depend on $Re_\lambda$. However, at a given $Re_\lambda$, the size of the intermediate range is expected to decrease with increasing $r_0$, since the Batchelor regime extends to larger $t/\tau_\eta$. Alternatively, the intermediate range can be more loosely defined as the full range between the end of the Batchelor regime and the onset of the Taylor regime. The latter definition is adopted when further quantifying the Reynolds number dependence of the intermediate range in §4.

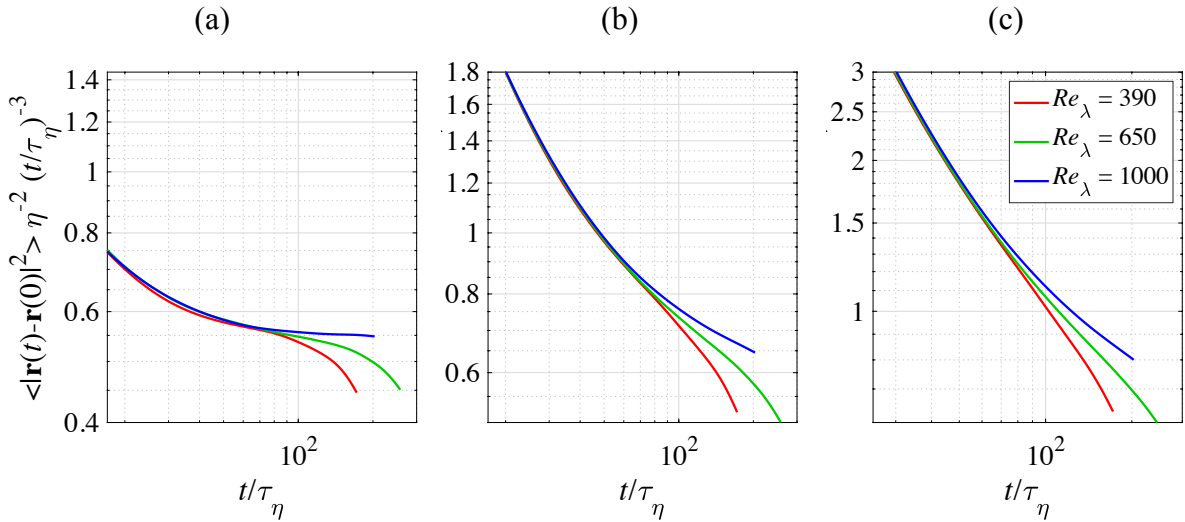

Figure 5. Reynolds number dependencies in the mean square separation for (a) $r_0 = 4\eta$, (b) $r_0 = 16\eta$ and (c) $r_0 = 64\eta$ (data from Buaria, Sawford & Yeung 2015). The curves present three different Reynolds numbers as indicated in (c).

Clear deviations from the inertial range dispersion model behavior (non-Richardson dispersion) have been reported for initial separations in the dissipative range, $r_0 < \eta$ (Scatamacchia, Biferale & Toschi 2012). However, some observations may still be of interest for the inertial range, because they illustrate how large-scales affect dispersion at smaller scales, including the smallest scale. The reported deviations were due to extreme events of rapidly separating pairs and pairs remaining close together for long time (order integral time scale). The separating velocity associated with the former is of the order of the rms velocity, $U$, which indicates a contribution from the large-scale energetic motions. So, both extremes are (partially) linked to the large scales, either through the time scale or the separation velocity. These large-scale influences at small $r_0$ can be understood from the shear layer structures that characterize the turbulent strain (Elsinga et al. 2017) and hence the turbulent dispersion. The velocity difference across the most intense small-scale straining structures within the shear layer is of order $U$ (Ishihara, Kaneda & Hunt 2013, Elsinga et al. 2017) meaning that a pair with $r_0 < \eta$ centered on such a structure separates with $U$. The pair then maintains that velocity difference for a significant amount of time (order integral time scale) as each particle enters into a different integral-scale flow region on either side of the shear layer (figure 1(a) pair A). This scenario is consistent with the extreme pair separation observed by Scatamacchia, Biferale

& Toschi (2012). Extremely slow dispersion on the other hand can be understood as pairs initially located within the same large-scale and nearly-uniform flow region bounding the shear layers (figure 1(a) pair B). Within these flow regions the velocity differences are relatively small, and because these regions are large, the particle pairs remain within them for long times (order integral time) without separating much. In these cases of extremely fast or extremely slow separation, the particles remember their initial condition (velocity difference) for very long times violating $t^3$ scaling law assumptions (§§2 and 3.1). Long term memory effects have also been reported by Bitane, Homann & Bec (2013) for $r_0 \leq 24\eta$. In these cases, the fact that the (initial) pair separation is small does not imply that there must be a small-scale eddy structure between these particles. In fact, if their relative velocity is small, it is more likely that they are in the same large-scale uniform flow region. Note that in the kinetic energy spectrum, low velocity magnitude is associated with small scales, but for relative dispersion the velocity difference, i.e. the velocity gradient, matters, whose magnitude is low at large scales. The above structural explanation for extreme dispersion highlights the important contributions of the large scales to the dispersion at small $r$, which is inconsistent with assumptions underlying the $t^3$ scaling law (§2).

*3.3 Non-classical pair dispersion in a significant shear layer*

Here, we present new results showing extreme pair separation. They are obtained for particles released inside the significant shear layer detected by Ishihara, Kaneda & Hunt (2013). The present results are based on the DNS of homogenous isotropic turbulence at $Re_\lambda = 1131$, where the full fluid velocity field has been advanced in time using the DNS code described by Ishihara et al. (2007, 2016). At the initial time, which corresponds to the time instant discussed in Ishihara, Kaneda & Hunt (2013), 719 fluid tracer particles are distributed randomly within 11 connected subdomains coinciding with the core of the detected significant shear layer. The subdomains are $96.4\eta \times 96.4\eta \times 96.4\eta$ in size and they contain the highest box-average enstrophy levels within a slice through the significant shear layer. The particles are tracked using the method described by Ishihara et al. (2018), who traced fluid particles and inertial particles in a series of DNS of turbulent flow using cubic spline interpolation for the fluid velocity at the particle position and using a fourth-order Runge-Kutta method for time integration. Here, we are concerned only with the fluid particle traces.

The significant shear layer was observed to survive in visualizations of intense vorticity until at least $t/\tau_\eta = 60$, where $t = 0$ corresponds to the time when the particles were released. During this time, the average strength of the vortices inside the layer decreased slightly. After $t/\tau_\eta \approx 60$, layer deformations were observed. However, a layer-type structure appeared to be maintained between the same large-scale motions for up to at least $t/\tau_\eta \approx 100$. At this point, it is hard to provide an exact number for the layer's lifetime. For the present purpose, it suffices that the layer survives until at least $t/\tau_\eta = 60$, because the particles leave the layer well before. This is confirmed by the results presented below. As pointed out before, the lifetime of the individual small-scale vortices inside the layer may be shorter.

From the particles released inside the significant shear layer, 1909, 2696 and 1920 particle pairs are obtained at $r_0 = 64\eta$, $128\eta$ and $256\eta$ respectively. These initial separations are in the inertial range and smaller than the significant shear layer thickness, i.e. $4\lambda_T = 264\eta$ at the present Reynolds number. While the pairs at $r_0 = 64\eta$ and $128\eta$ are approximately randomly oriented, the pairs at $r_0 = 256\eta$ are predominantly aligned with the significant shear layer because of confinement effects, which are caused by the separation being very close to the layer thickness. Therefore, the results at $r_0 = 256\eta$ are affected by orientation.

The mean square separation for the particle pairs released inside the significant shear layer is shown in figure 6. As expected, the mean square separation initially follows Batchelor scaling, which corresponds to a horizontal line in this $t^2$ compensated plot. The associated separation velocity is around $\sqrt{\langle|r(t)-r(0)|^2\rangle/t^2} = 32u_\eta$, which corresponds to $1.8U$. The large, order $U$, separation velocity is consistent with the velocity difference across the significant shear layer (see Ishihara, Kaneda & Hunt (2013) and §1). Moreover, the separation velocity in the Batchelor regime is relatively insensitive to $r_0$ within the considered range, which is clearly different from the unconditional result for pairs released anywhere in the flow at the same $r_0$ and similar $Re_\lambda$ (figure 4, blue lines). This also suggests that $U$ is the only relevant velocity scale for the dispersion inside the significant shear layer.

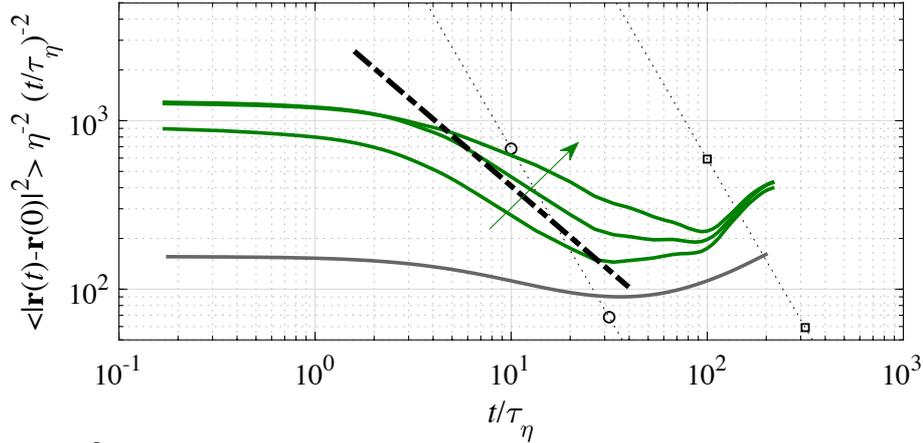

Figure 6. The $t^2$ compensated mean square separation for pairs released inside the significant shear layer (green lines). Different lines correspond to initial separations $r_0/\eta = 64$, 128 and 256 (increasing in the direction of the arrow). The unconditional result for pair released anywhere in the flow with $r_0/\eta = 64$ (grey line) is included for reference (from figure 4, data: Buaria, Sawford & Yeung (2015)). The thick dash-dotted line shows equation (3.1). Dashed lines with symbols (circles) and (squares) mark the condition $\langle|r(t)-r(0)|^2\rangle = (4\lambda_T)^2$ and $\langle|r(t)-r(0)|^2\rangle = L^2$ respectively.

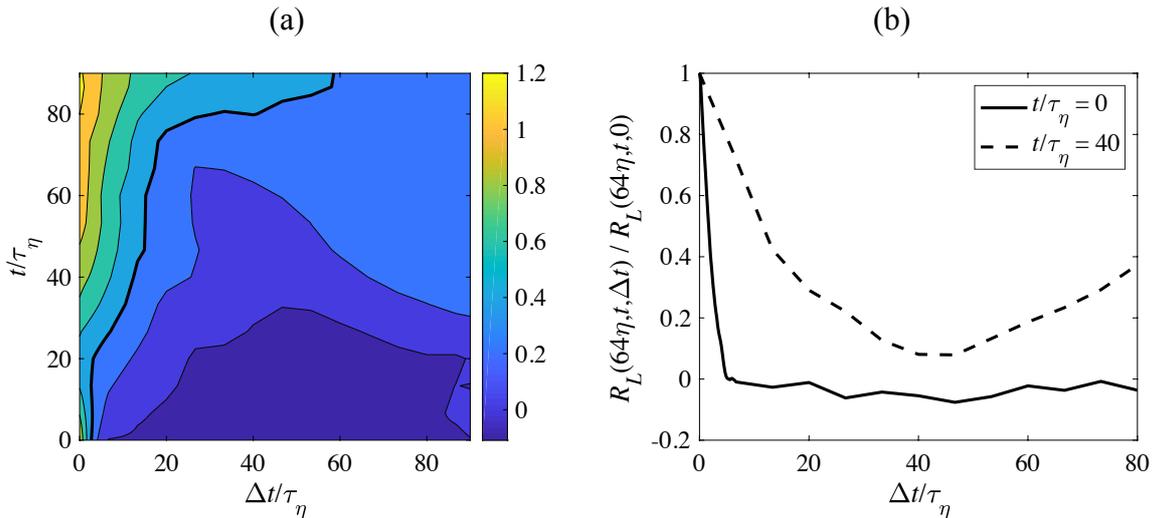

Figure 7. (a) Map of the Lagrangian correlation of the velocity difference, $R_L$, for $r_0 = 64\eta$ normalized by its value at $t = \Delta t = 0$. The thick contour indicates 0.4. (b) Profiles of $R_L$ taken at $t = 0$ and $t = 40\tau_\eta$.

In the range $4 < t/\tau_\eta < 20$, the mean square separation follows a $\sim t^{1.2}$ scaling for $r_0 = 64\eta$ and $128\eta$ (figure 6), which is reminiscent of a $t^1$ scaling associated with Taylor dispersion. However, it concerns a Taylor-like dispersion at relatively small scales, i.e. $\langle |\mathbf{r}(t) - \mathbf{r}(0)|^2 \rangle \lesssim (4\lambda_T)^2$, in contrast to the genuine Taylor regime, which appears at scales much larger than $L$. In Taylor dispersion the particles move independently, which is also the case inside the significant shear layer. This can be seen from the Lagrangian correlation of the velocity difference $R_L(r_0, t, \Delta t) \equiv \langle \delta \mathbf{u}(r_0, t) \cdot \delta \mathbf{u}(r_0, t + \Delta t) \rangle$, where $\delta \mathbf{u}$ is the relative velocity between the particles with the initial separation $r_0$. For $r_0 = 64\eta$ and $t/\tau_\eta < 20$, the normalized $R_L$ drops to small values (<0.4) within a time delay $\Delta t$ of $5\tau_\eta$ (figure 7(b)), which means that the relative velocity decorrelates quickly and that the particle motions are nearly independent after $5\tau_\eta$. This time scale is small, but not negligible, on the temporal range $4 < t/\tau_\eta < 20$. Moreover, the particle separation is still close to the small-scale coherence length and depends on $r_0$. Additionally, Taylor dispersion requires $R_L(r_0,t,0)$ to be constant, which is approximately the case (figure 7). Therefore, the dispersion is Taylor-like, but not exactly Taylor. The largely independent motion of the particles is consistent with the structural description of the significant shear layer, where intense small-scale structures are clustered inside a thin layer bounded by large-scale motions (§1). In the layer, there are no intermediate scales contributing importantly to the dispersion. The small-scale coherence length is between $60\eta$ and $120\eta$ depending on the definition (Elsinga et al. 2017, see also §3.1). Therefore, the Taylor-like dispersion by uncorrelated small-scale structures appears for separations larger than $60\eta$ and smaller than the layer thickness, $4\lambda_T$. Beyond $r \sim 4\lambda_T$, particles leave the layer and the bounding large-scale motions also contribute to their dispersion, which is therefore no longer Taylor-like. Consequently, the Taylor-like regime is present for $r_0 = 64\eta$ and $128\eta$, but is not as clearly seen for $r_0 = 256\eta = 3.9\lambda_T$.

Furthermore, the present observation of a Taylor-like regime appears consistent with the model analysis of Devenish & Thomson (2019). They showed that diffusion dominates pair dispersion after the initial Batchelor regime in case that (i) the constant of proportionality in the second-order Lagrangian velocity structure function (and hence the relative velocity) is very large and (ii) the relative velocity is short correlated in time as assumed in a Markov process description. As shown here, these conditions are satisfied locally within the significant shear layer.

The Taylor-like dispersion at intermediate $r_0$ can be quantified using the relation $\langle |\mathbf{r}(t) - \mathbf{r}(0)|^2 \rangle = 2 \int_0^t ds \int_{-s}^0 d\Delta s\, R_L(r_0, s, \Delta s)$ (Ishihara & Kaneda 2002). In the Taylor-like regime $R_L$ drops to zero quickly, which justifies introducing the approximation $\int_{-s}^0 d\Delta s\, R_L(r_0, s, \Delta s) \approx R_L(r_0, s, 0)\tau_L$, where $\tau_L$ is a Lagrangian correlation time scale for the small-scale motions inside the layer. Furthermore, $R_L(r_0, s, 0)$ is approximately constant in the Taylor-like regime. For $r_0 = 64\eta$ and $4 < t/\tau_\eta < 20$, our results suggest that $\tau_L \approx 4\tau_\eta$ and $R_L \approx 1.8U^2$, which yields:

$$\langle |\mathbf{r}(t) - \mathbf{r}(0)|^2 \rangle \approx 14 U^2 \tau_\eta t \qquad (3.1)$$

for the Taylor-like dispersion regime inside the significant shear layer. When compared to the data, relation (3.1) is seen to capture the order of magnitude in the Taylor-like dispersion range $4 < t/\tau_\eta < 20$ (figure 6).

For $r_0 = 64\eta$ and $128\eta$, other regime changes are observed at $t/\tau_\eta \approx 30$ and $100$ when $\langle |\mathbf{r}(t) - \mathbf{r}(0)|^2 \rangle$ is larger than $(4\lambda_T)^2$ and is approaching $L^2$ respectively (figure 6). The former is associated with pairs leaving the significant shear layer and entering the large-scale motions bounding the layer. The latter is close to the large-scale turnover time, $\sim 130\tau_\eta$ at the present Reynolds number, which is the order of magnitude for the lifetime of the significant shear layers (§1). Therefore, the particles have left the shear layer ($t/\tau_\eta \approx 30$) well before the expected lifetime of the shear layer ($>60\tau_\eta$). Just before the regime changes in figure 6, the peak of the Lagrangian correlation $R_L$ is observed to broaden in the direction of $\Delta t$, which is evident from the thick contour line breaking away from the vertical axis at $t/\tau_\eta \approx 20$ and 75 in figure 7(a). These rather sudden changes in the peak width signify that distinctly different scales affect the dispersion, which is very different from a gradual local dispersion process where the particles continuously probe the scale given by their instantaneous separation $r$. Furthermore, a longer-time velocity correlation appears for $t/\tau_\eta > 30$, when $R_L$ increases with $\Delta t$. Though eventually $R_L$ is expected to go to zero at very large $\Delta t$. This larger-scale influence in the correlation is consistent with particles entering into the large scale, quasi uniform flow regions bounding the significant shear layer at $t/\tau_\eta \approx 30$. Beyond $t/\tau_\eta \approx 100$, the mean square separation increases rapidly, but this result cannot be evidence for a Richardson scaling regime, because the temporal range is too short and the average separation is partially outside the inertial range (see criteria in §3.1).

Based on these new results, it is concluded that pair dispersion inside the significant shear layer is highly non-classical, because it is dominated by non-local dispersion. This non-locality is evident from (i) the velocity scale governing the dispersion, which is $U$ even at relatively small spatial scales, and (ii) the Taylor-like dispersion driven by small-scale motions, which appears for intermediate $r_0$. The latter is very different from the Batchelor and Richardson regimes commonly associated with small and intermediate time dispersion. Furthermore, we speculate that the Taylor-like dispersion inside the significant shear layers contributes to (or possibly causes) the dip in the unconditional $t^2$ compensated mean square separation (§3.2, figure 4), because they occur over the same temporal range.

The question remains, how important is the significant shear layer dispersion to the overall (unconditional) mean square separation? Its relative contribution is given by the product of the relative magnitude of $\langle |\boldsymbol{r}(t) - \boldsymbol{r}(0)|^2 \rangle$ and the volume fraction occupied by significant shear layers, assuming that the particle pairs are randomly distributed over space. The volume fraction occupied by significant shear layers is estimated to be ~10% at $Re_\lambda \approx 1000$ (Elsinga, Ishihara & Hunt 2020). Initially, the mean square separation inside the layer is almost an order of magnitude larger than the unconditional result for $r_0 = 64\eta$ and $Re_\lambda \approx 1000$ (see figure 6). At later times, the difference is still at least a factor 2–3. Multiplying these factors by the 10% volume fraction reveals that the significant shear layers' relative contribution to the overall mean square separation is significant and order one. Moreover, the significant shear layers will also contribute to the dispersion of the particles initially outside the layer and later entering into the layer, which is not accounted for here. Therefore, the present estimate represents a lower bound for their relative contribution.

The significant shear layer contribution remains significant as the Reynolds number increases. The volume fraction of these layers decreases according to $(\lambda_T L^2/L^3) \sim Re_\lambda^{-1}$ (Elsinga, Ishihara & Hunt 2020), while the layer's relative magnitude of $\langle |\boldsymbol{r}(t) - \boldsymbol{r}(0)|^2 \rangle$ scales according to $(U/u_\eta)^2 \sim Re_\lambda^1$, since the dispersion inside the layer and the overall dispersion are governed by $U$ and $u_\eta$ respectively. Therefore, the product of these two is independent of $Re_\lambda$ meaning that the layer's relative contribution to the unconditional mean square separation is order one at all Reynolds numbers.

*3.4 Results from kinematic simulations*

Richardson scaling has also been examined using kinematic simulations, where the particles are tracked through a random velocity fields with the spectral properties of a turbulent flow. However, as argued by Thomson & Devenish (2005), any appearance of a $t^3$ scaling range is accidental and the Richardson constants thus obtained are generally much smaller than those from experiments and numerical simulations of actual turbulence (e.g. Ott & Mann 2000, Sawford, Yeung & Hackl 2008). Their main explanation for the disparity is that, in kinematic simulations, the small-scales are not advected by the large scales as in actual turbulence, resulting in the particle motion being affected by the small-scales for too short times. However, Fung et al. (1992) in their kinematic simulations compensate for large-scale advection effects. Such compensation is non-trivial and introduces assumptions. These assumptions are essentially the same as those used in the theories discussed in §2, that is, the large-scales (larger than $r$) do not contribute and hence the dispersion is governed by eddies at scales close to the particle separation, $r$. Moreover, the phases of the different wavelengths are random in kinematic simulations, which is similar to the theories in the sense that the scales are considered to be independent. Therefore, it is not surprising that the kinematic simulations with large-scale advection compensation yield the same $t^3$ scaling law as the theories (e.g. Fung et al. 1992). It proves that the theory is internally consistent, but it cannot validate the theory, because similar simplifying assumptions are used. It is, in fact, these assumptions that are being challenged by the lack of an observable Richardson scaling regime in DNS data (§3.2) and by the results in §3.3 showing that the large scales are significantly contributing to the dispersion at $r \ll L$.

As mentioned, the large- and small-scale modes are independent in kinematic simulations by construction. This leads to a profoundly different turbulent flow structure, which affects the particle dispersion. It was already observed by Fung et al. (1992) that the length of the intense vorticity tubes was underestimated by their artificial velocity fields. The same effect is visible in the visualizations of the intense vortices and the large-scale uniform flow regions in Elsinga & Marusic (2010), their figures 14 and 15. Not only are the vortices much shorter, also the shape of the large-scale structure and the spatial distribution of the small scales are different. In actual turbulence, the large-scales appear elongated and the intense small-scale vortices form clusters along their edges. However, in the artificial field, the large scales are blob-like and the small-scales are distributed uniformly over the entire domain. The dispersion of particle pairs in these different velocity fields is illustrated in figure 8. For the purpose of particle tracking both fields were treated as stationary. After the initial Batchelor regime ($t/\tau_\eta > 2$), the scaling exponents (i.e. the slopes in figure 8) are observed to differ, especially for the larger initial separations $r_0/\eta = 65$ and 130. The coherence of the velocity field in the actual DNS is found to slow down the pair dispersion as compared to the random field. Note that this illustration concerns a low Reynolds number. When significant shear layers appear ($Re_\lambda > 250$), the spatial clustering of intense small-scale structures, i.e. intermittency, is much more pronounced, increasing the differences with respect to a random velocity field.

To summarize, kinematic simulations with large-scale advection compensation cannot be used to validate the Richardson scaling law. Moreover, the artificial velocity fields do not accurately reflect the turbulent flow structure and the pair dispersion. Because of these issues, we will not consider the kinematic simulations in further detail here.

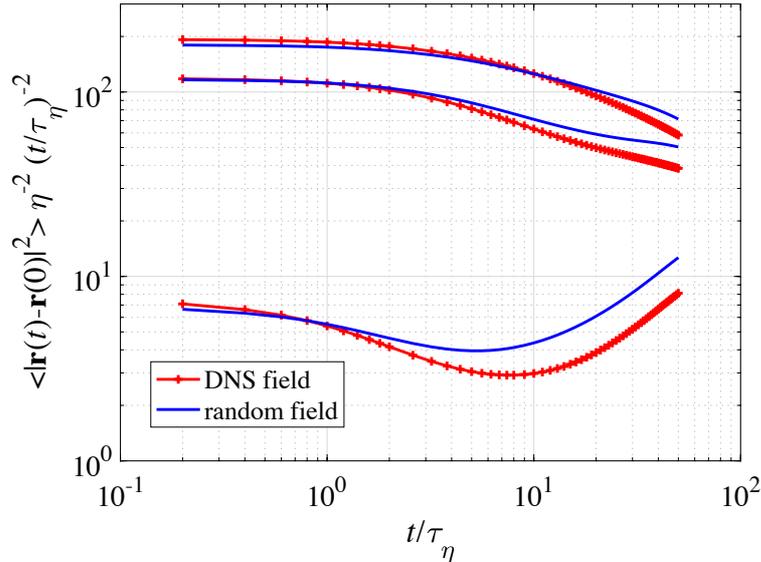

Figure 8. Kinematic simulations of pair dispersion in a random and DNS velocity field with the same turbulent kinetic energy spectrum ($Re_\lambda = 170$). The evolution of the $t^2$-compensated mean square separation is shown for three initial separations, $r_0/\eta = 5, 65$ and $130$ (increasing upwards). The DNS field was provided by Dr. A.A. Wray (CTR 2002, private communication). The random field was generated using the method of Rogallo (1981).

### *3.5 Results from experiments*

Unfortunately, laboratory experiments have advanced very little over the last decade as far as pair dispersion is concerned. The measurements of Ouellette et al. (2006) and Bourgoin et al. (2006) still represent the highest Reynolds numbers achieved in an experiment, i.e. $Re_\lambda = 815$. These flow conditions have since been surpassed by DNS reaching $Re_\lambda = 1000$ (§3.2). However, the mentioned experimental studies presented data for a wide range of initial separations within the inertial range up to approximately $t = 50\tau_\eta$, after which the scatter in the data increases. Only a Batchelor $t^2$ scaling regime was observed, which has later been attributed to limited observation time ($<t_t$, Bourgoin 2015). Also, it has been suggested that the finite measurement volume may play a role in these experiments (Salazar & Collins 2009). However, the experimental results appear generally consistent with DNS, which also showed that for $t < t_t$ the pair separation follows $t^\beta$ scaling with $\beta \leq 2$ (§3.2).

For low Reynolds number, $Re_\lambda \approx 100$, Ott & Mann (2000) showed a short $t^3$ scaling range when plotting their data using a virtual origin. This range extended over less than half a decade in time, which is not yet convincing for reasons explained in §3.1. Moreover, the $t^3$ scaling is observed over a temporal range that partially overlaps with the Batchelor scaling range, i.e. it extends below $t = t_B$, suggesting that $t^2$ and $t^3$ scaling are valid simultaneously. Furthermore, their results are for an initial separation outside inertial range, i.e. $r_0 \approx 10\eta$. The short $t^3$ scaling range was also observed in the results from a large-scale DNS at $Re_\lambda = 283$ covering a wider range of initial separations outside the inertial range, $r_0 \leq 32\eta$, but a virtual origin had to be used (Ishihara & Kaneda 2002).

Based on their experiments at $Re_\lambda \approx 180$, Shnapp & Liberzon (2018) found that pairs with low initial separation velocity magnitude, i.e. small $dr/dt$, separated according to a $t^3$ law, whereas pairs with high initial separation velocity separated following a $t^2$ scaling. However, these observations were made outside the inertial range, that is, $r_0 \leq 15\eta$ and for time scales smaller than the eddy turnover time at $r_0$. Nevertheless, it reinforces the notion that the mean square separation represents an average over widely different separation behaviors.

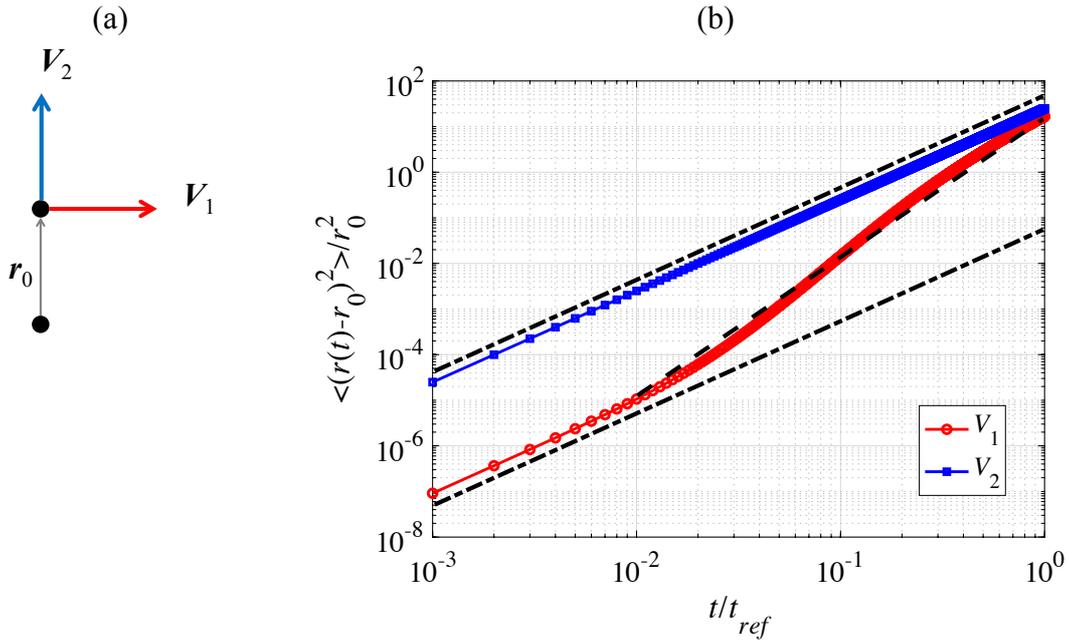

Figure 9. (a) A model to illustrate the effect of the orientation of the separation velocity $V$ relative to the initial pair separation vector $r_0$. The normal velocity $V_1$ and parallel velocity $V_2$ are taken with respect to the velocity of the bottom particle. (b) The resulting mean square separation for the two cases shown in (a) assuming constant velocity. The results have been averaged considering a small range of deviations in angles from the mean directions indicated in (a) (<6 degrees). The dash-dotted lines and dashed line indicate $t^2$ and $t^3$ power laws respectively.

The observation that a subset of pair dispersions follows a $t^3$ law is interesting, and remains to be explained. We suggest that it might be related to the orientation of the separation velocity vector relative to the initial separation vector $r_0$. Hereto, consider two tracer particles in a frame of reference attached to the bottom particle (figure 9(a)). In this frame, the velocity of the other particle, $V$, corresponds to the separation velocity vector. When $V$ is assumed constant and normal to $r_0$ ($V_1$ in figure 9(a)), then $dr/dt$ is initially zero, or small when allowing for a slight deviation from the normal condition. The resulting evolution of the mean square separation $\langle (r(t) - r(0))^2 \rangle$ is given in figure 9(b), where $r(t) = |r(t)|$. The plot shows that after the initial Batchelor $t^2$ regime, the separation is accelerated as $r$ tilts in the direction of $V_1$. During this stage, the dispersion closely follows a $t^3$ scaling law. At later times (beyond the scale of the plot) $t^2$ scaling is recovered when the tilting is complete and $r$ approximately aligns with $V_1$. On the other hand, if $V$ and $r_0$ align ($V_2$ in figure 9(a)), then the initial $dr/dt$ is significant and a $t^2$ scaling is found throughout (figure 9(b)). The present results are qualitatively consistent with those of Shnapp & Liberzon (2018), which suggests that the initial orientation of the relative velocity is important. We further comment that identical results are obtained for both orientations of $V$ when the vector form $\langle |r(t) - r(0)|^2 \rangle$ is used to quantify the mean square separation, instead of the scalar form $\langle (r(t) - r(0))^2 \rangle$ (figure 9(b) and Shnapp & Liberzon 2018). Therefore, also the metric used to quantify the mean square separation is important.

Many of the more recent experiments have considered inertial particles, and the effects of walls or thermal convection, which is different from the present problem of fluid tracers in homogeneous isotropic turbulence. But also in these cases, $Re_\lambda$ is typically in the range of 250-500 (e.g. Dou et al. (2018), Petersen Baker & Coletti (2019)), which is limiting when looking for a Richardson scaling regime. However, there is some evidence to suggest that also in Rayleigh-Bénard convection a $t^3$ scaling range appears for $r_0 \approx 4\eta$ with other $r_0$ approaching from above or below (Liot et al. 2019). Pair dispersion experiments at much higher Reynolds numbers would be of considerable interest. New facilities can generate grid turbulence at $Re_\lambda$ ~5000 (Küchler, Bewley & Bodenschatz 2019), albeit grid turbulence is decaying, but they have not yet been used to measure pair dispersion.

High Reynolds numbers are also achieved in environmental flows, but the experimental conditions are typically poorly controlled and the flow is often inhomogeneous and affected by buoyancy (Salazar & Collins 2009), which introduces considerable uncertainty. Therefore, these experiments are not considered here.

*3.6 Summary*

At present, there is no clear and indisputable evidence for a Richardson $t^3$ scaling regime in the inertial range. Only for the case of $r_0 \approx 4\eta$ did a $t^3$ scaling range appear, which could be coincidental. A genuine Richardson regime would have resulted in $t^3$ scaling for a range of initial separations within the inertial range ($r_0 > 60\eta$), not just for one specific case of $r_0 \approx 4\eta$ outside this range. Other initial separations were found to approach the $r_0 \approx 4\eta$ result from above or below, but they never reached a true $t^3$ scaling regime.

Despite a considerable increase in the Reynolds numbers achieved by DNS, the separation of scales remains limited. At $Re_\lambda = 1000$, the mean separation is of the order of the significant shear layer thickness, i.e. $\langle |\boldsymbol{r}(t) - \boldsymbol{r}(0)|^2 \rangle \approx (4\lambda_T)^2$, at the end of the initial Batchelor regime. It means that some pairs are dispersed by large-scale motions well before a Richardson regime might develop (e.g. figure 1(a) pair A).

New results for pairs released inside a significant shear layer confirmed this conjecture. They revealed that the separation velocity is of the order of $U$, and that the dispersion is Taylor-like for $r_0 > 60\eta$ and $4 < t/\tau_\eta < 20$. This suggests that the large scales contribute to the dispersion through the velocity scale, while the independent small-scale motions inside the layer govern the particle motions. There does not seem to be an important role for intermediate scales, which strongly questions the validity of the assumptions leading to the prediction of a Richardson scaling regime. Furthermore, scaling analysis showed that the non-local contribution from the significant shear layers to the overall mean square separation is significant at all Reynolds numbers.

Finally, the metric used to quantify the mean square separation can have important consequences for the observed dispersion behavior. For example, the introduction of a virtual origin may introduce a spurious $t^3$ scaling range over approximately half a decade of time. Also, there can be profound differences between the vector and the scalar definitions of the mean square separation. These issues sometimes complicate a straightforward comparison of results. However, they do not explain the fact that a Richardson scaling regime has not been observed in the studies discussed.

**4. The intermediate range as a transition region**

Here, we propose an alternative view of pair dispersion in the intermediate range, which is consistent with the available data. Starting from the well-established initial Batchelor regime and the ultimate Taylor regime, the intermediate range is regarded simply as a transition between these two regimes. This is similar to considering the intermediate range in the velocity structure functions as a matching region (§2.4). Because we do not use inertial range assumptions, the results of our model apply to wide range of $r_0$, not limited to the inertial range. Also, the present model is consistent with non-local contributions to the dispersion.

The initial pair separation is ballistic meaning that the particles maintain their initial velocity for sufficiently short time-scales. This may be relaxed by assuming that the mean square separation velocity, $\langle|\delta\boldsymbol{u}(r,t)|^2\rangle$, is constant, which is valid for longer times (e.g. Goudar & Elsinga 2018) and leads to the so-called Batchelor scaling regime (Batchelor 1950):

$$\langle|\boldsymbol{r}(t) - \boldsymbol{r}(0)|^2\rangle = \langle|\delta\boldsymbol{u}(r_0,0)|^2\rangle t^2 = S_2(r_0) t^2 \tag{4.1}$$

where $\delta\boldsymbol{u}$ is the relative velocity between the particles. To compute the three-dimensional second-order structure function, $S_2$, we use the functional form given in Donzis & Sreenivasan (2010), which is given by:

$$S_2(r) = 3f(r) + r\frac{df(r)}{dr}$$

$$\text{with} \quad f(r) = \tfrac{1}{15} u_\eta^2 \frac{\left(\frac{r}{\eta}\right)^2}{\left[1+\left(c_B\frac{r}{\eta}\right)^2\right]^{2/3}} \tag{4.2}$$

where $c_B = 0.076$. The Batchelor regime extends up to $t = 3t_B$ in our model. Note that the Batchelor regime is Reynolds number independent in Kolmogorov scaling. Therefore, it seems reasonable to assume that the end of this regime scales accordingly, and hence is proportional to $t_B$ independent of the Reynolds number. The pre-factor is empirical and estimated from figure 4, where we have included the part revealing a slight dip into the Batchelor regime. Moreover, $3t_B$ is a close approximation of the time scale $t_t$ proposed by Bitane, Homann & Bec (2013) for the end of the Batchelor regime, when $r_0 > 60\eta$ and using (4.2).

At large time, the pair separation is of order $L$ or larger, which implies that the motions of the particles are uncorrelated and independent of the initial separation $r_0$. This is known as the Taylor regime, in which the mean square separation grows as (Taylor 1922, Pope 2000, Buaria, Sawford & Yeung 2015):

$$\langle|\boldsymbol{r}(t) - \boldsymbol{r}(0)|^2\rangle = 12 U^2 T_L t \tag{4.3}$$

Define the onset of the Taylor regime at $t = cT_L$, where $c \approx 8$ is a constant independent of $r_0$ and the Reynolds number. Here, the simplifying assumption is that the Taylor regime starts at the same value of the mean square separation, and hence time, for all $r_0$. This seems reasonable given the convergence observed in figure 4 (and also in figure 11(b)). Furthermore, the onset of the Taylor regime is assumed to collapse when using integral scales, which is consistent with the general scaling in this regime. The value of $c$ was estimated from the intersection point when extrapolating the data for different $r_0$ (see figure 4). Furthermore, using $T_L = 2U^2/(D_L \varepsilon)$ (Tennekes 1979, Pope 2000, Sawford, Yeung & Hackl 2008), we obtain for the onset of the Taylor regime:

$$\frac{\langle|\boldsymbol{r}(cT_L) - \boldsymbol{r}(0)|^2\rangle}{\eta^2} = \frac{6cD_L \varepsilon T_L^3}{\eta^2} = 6cD_L\left(\frac{T_L}{\tau_\eta}\right)^3 \tag{4.4}$$

The expression used for $T_L$ can be understood from the energy balance, where the average production-rate scales with large-scale quantities (i.e. $U^2/T_L$) apart from a constant and is equal to the average dissipation-rate $\varepsilon$ (see also §2.4). Lacking a better estimate, the value of $D_L$ has been determined from the inertial range of the second-order Lagrangian structure function as $D_L \approx C_0$, where $C_0 = 6.0$ is the Lagrangian Kolmogorov constant (Lien & D'Asaro 2002, Sawford, Yeung & Hackl 2008, Sawford & Yeung 2011). However, this inertial range approach is considered reasonable for second-order moments of velocity as explained in §2.4. Rewriting equation (4.4) in a time compensated form gives:

$$\frac{\langle |\mathbf{r}(cT_L) - \mathbf{r}(0)|^2 \rangle}{\eta^2} \left(\frac{\tau_\eta}{cT_L}\right)^3 = \frac{6D_L}{c^2} = g' \qquad (4.5)$$

which yields a constant $g' \approx 0.56$ independent of $r_0$ and the Reynolds number. This means that the point marking the onset of the Taylor regime shifts along the time axis according to $cT_L/\tau_\eta$ when changing the Reynolds number, but remains constant in terms of the $t^3$ compensated mean square separation.

Finally, the intermediate regime is coarsely approximated by a basic $t^\beta$ power law connecting the end point of the Batchelor regime with the starting point of the Taylor regime (figure 10). In the logarithmic plot, this results in a straight-line connection, where the slope determines the exponent $\beta$. It represents a strong simplification, since the actual Batchelor-to-intermediate and intermediate-to-Taylor regime transitions are smooth. Nevertheless, the simple model allows to predict the exponent of the intermediate power law as a function of the Reynolds number and the initial separation, which can be compared to actual data.

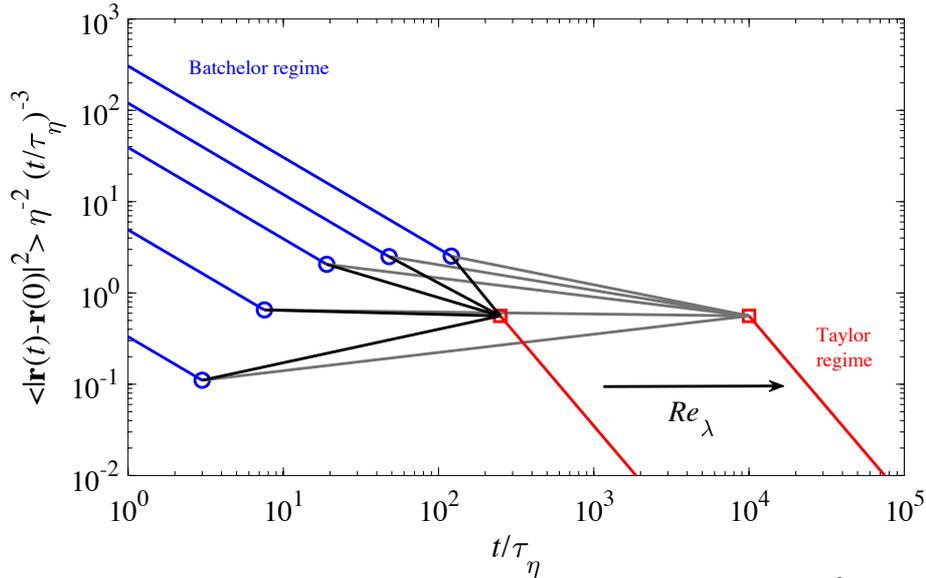

Figure 10. Model for the mean square pair separation presented in the $t^3$ compensated plot. The initial Batchelor regime (equation (4.1), blue lines) is shown for different initial separations, $r_0/\eta = 1, 4, 16, 64, 256$. The end time of this regime is $3t_B$ (circles). The final Taylor regime (equation (4.3), red lines) is shown for two different Reynolds numbers. The onset of the Taylor regime (squares) is at $t = cT_L$ along the horizontal axis, while it is constant along the vertical axis due to the chosen normalization, see text, and independent of $r_0$. In this model, the intermediate range is represented by a power scaling, which connects the endpoint of the Batchelor regime with the starting point of the Taylor regime (black and grey lines corresponding to different Reynolds numbers). The latter is a simplification, because in reality the regime changes are smooth.

The present model (figure 10) illustrates how the mean square separation slowly approaches $t^3$ scaling in the limit of $Re_\lambda \to \infty$, but never reaches it in any real flow. Note that $t^3$ scaling corresponds to a horizontal line in this $t^3$ compensated plot. Furthermore, it is worth pointing out that the high Reynolds number case shown in figure 10 is representative of atmospheric conditions with $Re_\lambda \sim 10^4$, where the onset of the Taylor regime is anticipated at $cT_L/\tau_\eta \sim 10^4$. So, even in these highly turbulent flows, Richardson scaling is not returned by the model. There is, however, one exception. For $r_0/\eta \approx 4$, the end of the Batchelor regime and the start of the Taylor regime are on the same vertical coordinate, which results in $t^3$ scaling. This is consistent with DNS revealing $t^3$ scaling only for $r_0/\eta \approx 4$ (Bitane, Homann & Bec 2013, Bragg, Ireland & Collins 2016, Buaria, Sawford & Yeung 2015, see also §3.2). Moreover, as noted by Sawford, Yeung & Hackl (2008), the compensated relative dispersion will approach the $t^3$ scaling from below for $r_0/\eta \lesssim 4$ and from above for $r_0/\eta \gtrsim 4$, which is reproduced by the model.

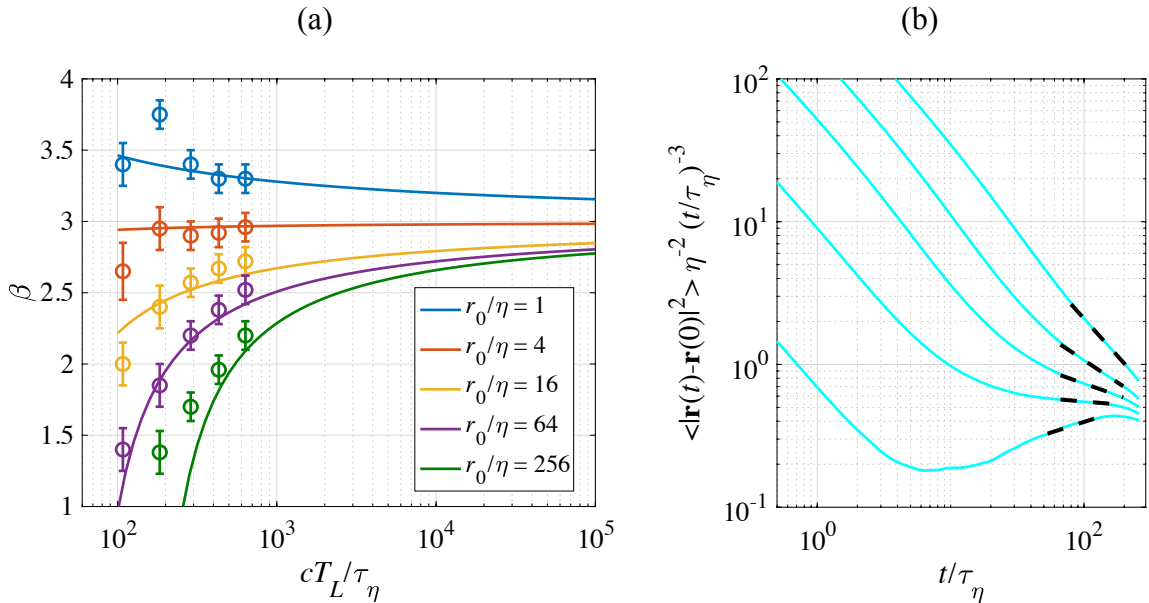

Figure 11. (a) The $t^\beta$ power law exponent for the intermediate regime versus $cT_L/\tau_\eta$, which scales linearly with $Re_\lambda$. Solid lines show the model prediction for different initial separations, $r_0$, where $\beta$ has been determined from the power law connecting the Batchelor and Taylor regimes (black and grey lines in figure 10). Circles present results from power law fits to DNS pair dispersion data at $Re_\lambda$ = 140, 240, 390, 650 and 1000 (data sources: Sawford, Yeung & Hackl 2008 and Buaria, Sawford & Yeung 2015), as illustrated in (b). The error bars indicate the estimated accuracy of the fit. (b) Power laws fits (dashed lines) to the intermediate regime of the mean square separation for the case $Re_\lambda = 650$ (solid lines, source: Buaria, Sawford & Yeung 2015).

The rate at which $t^3$ scaling is approached is presented in figure 11(a) for different $r_0$. The plot shows the power law exponent $\beta$ versus the time corresponding to the onset of the Taylor regime, $cT_L/\tau_\eta$, which varies linearly with $Re_\lambda$. The results in figure 11(a) confirm our discussion of figure 10 by showing that $\beta = 3$ for $r_0/\eta \approx 4$, while, at finite $Re_\lambda$, larger and smaller $r_0$ yield $\beta < 3$ and $\beta > 3$ respectively. Only in the limit of $Re_\lambda \to \infty$ is $\beta = 3$ independent of $r_0$. Furthermore, figure 11(a) compares the model predictions (lines) with power law fits to actual DNS data (circles). These fits were obtained in the middle of the intermediate regime ($3t_B < t < cT_L$) where the slope of the mean square separation was approximately constant in the logarithmic plot. These fits are illustrated in figure 11(b) for the case of $Re_\lambda = 650$. Apart from the lowest Reynolds number, the model agrees with the actual DNS data to within the accuracy of the fit. The deviations at $Re_\lambda = 140$ ($cT_L/\tau_\eta = 110$) may be understood from the fact that the straining motions are underdeveloped for $Re_\lambda < 250$ (Elsinga et al. 2017), which affects the dispersion. At $Re_\lambda = 240$ ($cT_L/\tau_\eta = 180$), $r_0/\eta = 256$ corresponds to $r_0 \approx L$, which means that $r_0$ is not in the intermediate-$r$ range. In that case, it is possible that there is no intermediate regime and the Taylor regime directly follows the Batchelor regime. This explains the deviation of this data point from the model prediction. All remaining data points, and especially those within the intermediate-$r$ range ($r_0/\eta = 64$), seem consistent with the model.

While the model results slowly approach $t^3$ scaling with increasing Reynolds number (figure 11(a)), there is an important fundamental difference between the present model and the theories leading to the Richardson scaling law (§2). The theories assume local dispersion and invoke inertial range assumptions, where the dispersion is independent of the large (energetic) scales as well as the dissipative scales. By contrast, the present model uses the properties of the large scales, i.e. $U$ and $L$, (equation 4.3) and the small time-scales, i.e. $u_\eta$, $\eta$ and $r_0$, (equation 4.1) and assumes a transition region. This introduces a $r_0$ and Reynolds number dependence in the scaling exponent for the intermediate range. These dependencies are observed in the DNS data (§3.2), but are not retained in K41-based theory as discussed in §2.4. Basically, the present model yields an overlap region for each $r_0$ separately. Further note that the classical results for $S_2$ and $C_0$ used here, can also be obtained without inertial range assumptions (see §2.4). The present model explains why $t^3$ scaling is found for just one specific initial condition outside the inertial range (i.e. $r_0/\eta \approx 4$ independent of $Re_\lambda$), and also captures the departures for $r_0/\eta \lesssim 4$ and $r_0/\eta \gtrsim 4$, which the inertial theories (§2) do not. Furthermore, the 'Richardson constant' is $g' \approx 0.56$ for $r_0/\eta \approx 4$ (equation (4.5)), which is consistent with the reported values in the literature ($g = 0.55 – 0.57$, Sawford, Yeung & Hackl 2008). We should point out that the value of $r_0$ for exact $t^3$ scaling and the value of $g'$ are not predicted from first principles. They partially depend on observations related to the starting point and the end point of the intermediate range. Therefore, consistency between the resulting $g'$ and the reported $g$ is expected, even if it is not enforced. Furthermore, $g'$, and to a lesser extent $r_0$ for which $t^3$ scaling is obtained, is sensitive to the uncertainties in $C_0$ (Lien & D'Asaro 2002) and $c$, see equation (4.5). Nevertheless, the present estimate for the onset of the Taylor regime, i.e. $c \approx 8$, seems to return a reasonable value for $g'$.

The model introduces two empirical coefficients, i.e. $c$ and the pre-factor associated with $t_B$, which means that at most two of the model outcomes can be fixed by these coefficients. For sake of argument, let's say the fixed outcomes are the value of $g'$ and the value of $r_0$ for exact $t^3$ scaling. The former depends on the coefficient $c$ and the latter depends on both coefficients. Once the coefficients are set, the exponents for the other $r_0$ and their Reynolds number dependencies (figure 11) follow directly from the model and cannot be adjusted. Therefore, the most significant results from the model are (i) the Reynolds number dependence of the intermediate range for different $r_0$ and (ii) the finding that there is only a single $r_0$ for which exact $t^3$ scaling is obtained.

Only in the limit of $L/r_0 \to \infty$ and $L/\eta \to \infty$ (at $Re_\lambda \to \infty$) can a $r_0$ dependence be neglected in the present model, because $r_0$ and $\eta$ become indistinct. In that case, a Kolmogorov scaling regime at small $t$, independent of $r_0$, is matched to an integral scaling regime at large $t$, which yields the Richardson scaling consistent with K41, but without having to assume local dispersion within an inertial range where all flow properties depend only on $\varepsilon$ and $r$.

Presently, a power law has been assumed for the intermediate range, and details, such as the dip in the $t^2$ compensated mean square separation (figure 4) and the smooth connections to the Batchelor and Taylor regimes, are ignored. A possible alternative to the power law assumption could be a (slightly) curved line connecting the Batchelor and Taylor regimes in figure 10.

Further work characterizing the shape of the connection is clearly warranted. However, including this kind of detail does not fundamentally alter the physical picture presented here, where the intermediate range is a transition region. It will yield a similar slow $r_0$-dependent approach of $t^3$ scaling when it is being stretched due to an increasing Reynolds number, i.e. increasing separation between Batchelor and Taylor regimes.

As commented before, dispersion at intermediate times is highly complex, because $r$ is widely distributed. Therefore, the pair dispersion is affected by many scales, or perhaps even the full range of scales, at any one time. The present model for the mean square separation incorporates some of these effects by considering the intermediate range as a transition region, where the balance gradually shifts from a state governed by $r_0$ and Kolmogorov scaling (i.e. Batchelor regime) to a state fully described by the large, energetic scales (i.e. Taylor regime). This transition scenario is consistent with the existence of significant shear layer structures, where only small and large scales contribute to the dispersion (figure 1(a) and §3.3). At intermediate times, we expect that a blend of these two contributions will determine the statistics.

Recently, Liot et al. (2019) have shown that the ballistic approach of Bourgoin (2015) (§2.2) can yield a $r_0$-dependent intermediate range consistent with observations and the present model. A $t^3$ scaling was obtained for $r_0/\eta \approx 4$, while other initial separations approached $t^3$ scaling from above or below. This was achieved by replacing the inertial range model $S_2 \sim (\varepsilon r)^{2/3}$ with an actual $S_2$, which included large- and small-scale ranges. It seems to confirm that all scales play a role in the intermediate range. Their results are for a single Reynolds number, $Re_\lambda \approx 75$, and $r_0/\eta \leq 10$. It would be of interest to extend them to larger $r_0$ and to higher Reynolds numbers. The Lagrangian Stochastic Models of Borgas & Yeung (2004) and Devenish & Thomson (2019) also yields a $t^3$ scaling for $r_0/\eta \approx 4$, with larger $r_0$ approaching from above, when using the actual probability distribution of the separation velocity instead of an inertial range model. This distribution was obtained directly from DNS at $Re_\lambda = 390$ (Devenish & Thomson 2019) or modelled after DNS results at $Re_\lambda = 90 - 230$ (Borgas & Yeung 2004) and included dissipation range effects and longer tails. The latter are associated with extreme separation velocities (intermittency) and non-local effects. However, model parameters had to be adjusted for each Reynolds number in order to obtain a good match between the model and the DNS in terms of the mean square separation and the value of $g$ at $r_0/\eta \approx 4$ (Borgas & Yeung 2004). This appears consistent with Sawford & Yeung (2010) showing that $g$ strongly increases with $Re_\lambda$ when using constant model parameters. The Reynolds number dependence of the model parameters remains to be explained. Therefore, the ballistic and Lagrangian stochastic modelling frameworks can give a realistic $r_0$ dependence in the intermediate range if non-inertial range behavior is included. It supports our assertion that the dispersion in the intermediate range should not be understood as inertial and local. The intermediate range is perhaps best seen as a transition region where non-local dispersion is important. A significant example of such non-local dispersion is given by the large-scale shear layers (§3.3).

## 5. Conclusions

The Richardson scaling law for the mean square separation of a tracer pair has been critically assessed. The common theories predicting this $t^3$ scaling regime assume that only the eddies at the scale of the (mean) pair separation contribute to the dispersion. This is referred to as local dispersion. Moreover, it is assumed that the pair separation is within an inertial range, where the pair separation is not affected by the small dissipative scales and the large energetic scales. It is believed that the increasing scale separation with increasing Reynolds number allows for such an inertial range to eventually develop. However, the assumption ignores any phase dependence between the scales, that is, any structural organization of the turbulent flow. Particularly relevant in this context are the significant shear layers, which develop at high Reynolds numbers. As shown in §3.3, particle pairs within these layers are initially transported by the small scales, but when their separation increases over time, they leave the small-scale structures and almost immediately enter into the large energetic flow regions bounding the layer (as illustrated for example by pair A in figure 1(a)). An inertial range contribution is, therefore, absent around the significant shear layer. This results in a Taylor-like dispersion inside these layers, which is very different from the classical dispersion regimes. Importantly, these layers provide a significant, and non-local, contribution to the overall mean square separation at all Reynolds numbers, because the particle separation velocity is of the order of $U$ inside the layer. Additionally, results reported in the literature show that the separation, $r$, is broadly distributed at intermediate times when the Richardson scaling law is expected to apply. It implies that some pairs have entered the small or the large scales (outside the inertial range). Consequently, all scales contribute to the dispersion at intermediate times, not just the intermediate / inertial scales. Furthermore, there is increasing evidence to suggest that even the dissipative scales contain important large-scale influences (§2.4), which would mean that the fluid motions, and hence tracer motions, are affected by the large energetic scales at all separations. These observations raise doubt on whether the assumption of local dispersion within an inertial range is justified.

Cases where a $t^3$ scaling has been observed do not support the theory for two main reasons. First, the $t^3$ scaling was observed over too short temporal ranges (less than a decade), which leaves open the possibility that the scaling is artificially introduced by the use of a virtual origin. Secondly, $t^3$ scaling was found exclusively for an initial separation $r_0/\eta \approx 4$, which is inconsistent with the theory. Specifically, the theory applies to $r$ (including $r_0$) in the inertial range and it predicts $t^3$ scaling for a range of $r_0$, as opposed to a single ('lucky') value for $r_0$ outside the inertial range. Extensions of the theory to $r_0 \lesssim \eta$ have been suggested in the literature, but this does not explain the available data either. Therefore, it is concluded that the classical theory relies on debatable assumptions and, so far, lacks evidence.

As an alternative to the current theories, we proposed in §4 that the intermediate regime is simply the transition region from the initial Batchelor regime to the ultimate Taylor regime. This is consistent with the observed non-local contribution to the dispersion, because the inputs contain only small-time-scale and large-scale turbulence properties. The simple transition model explains why there is only a single $r_0 \approx 4\eta$ for which there is a true $t^3$ scaling and why smaller/larger $r_0$ approaches $t^3$ scaling from below/above, as observed in actual data. The inertial range and local dispersion theories discussed in §2 do not explain this behavior. The 'Richardson constant' returned by the transition model is 0.56 for this $t^3$ regime at $r_0 \approx 4\eta$, which is consistent with the observed value in DNS. Furthermore, the model prediction of the power law exponent for the intermediate regime is consistent with the presently available dispersion data from DNS at finite Reynolds number. Only in the limit of infinite Reynolds number does the model yield a $t^3$ scaling regime for $r_0 \ll L$. In that case, the $t^3$ regime is an asymptotic state, which does allow for significant non-local contributions to the dispersion.

The intermediate range is relatively short within the presently available data ($Re_\lambda \lesssim 1000$), which introduces some uncertainty when extrapolating the results to higher Reynolds numbers. Accurate data at $Re_\lambda \sim 10^5$ seems desirable to probe the intermediate range in Lagrangian statistics (e.g. Lien & D'Asaro 2002 and Sawford & Yeung 2011). However, applications of turbulent dispersion are often found at lower $Re_\lambda$, which makes finite Reynolds number theories of important interest, in addition to the infinite Reynolds number limit.

**Acknowledgement**

We thank the referees for their comments, which helped to improve the manuscript. This work used computational resources of the K computer / the supercomputer Fugaku provided by RIKEN, and the Oakforest-PACS in the Information Technology Center, the university of Tokyo, through the HPCI System Research Project (Projects ID: hp200124 and hp210164). This work also used computational resources provided by the Information Technology Center, Nagoya University, and Research Institute for Information Technology, Kyushu University, through the HPCI Research Project (Projects ID: hp190084 and hp200042) and the JHPCN Joint Research Project. TI was supported in part by JSPS KAKENHI Grant Number 20H01948 and MEXT as "Program for Promoting Researches on the Supercomputer Fugaku" (Toward a unified view of the universe: from large scale structures to planets). We thank Dhawal Buaria and PK Yeung for providing their data at $Re_\lambda$ = 390, 650 and 1000, which was generated under NSF Grant No. CBET-1235906.

**Declaration of Interests.** The authors report no conflict of interest.

**Appendix A. Richardson scaling in linear plots**

Here, we briefly discuss the linear plot of $\langle r^2 \rangle^{1/3}$ versus time, where a Richardson scaling regime appears as a straight line. Such plots have been used in the past, for example by Ott & Mann (2000) and Ishihara & Kaneda (2002). It is shown that the condition for positively identifying a Richardson scaling regime is similar to that for the logarithmic plot.

Bitane, Homann & Bec (2013) provide $\langle r^2 \rangle$ at $Re_\lambda$ = 730. Figure 12 shows their result for $r_0 = 24\eta$, which is the largest initial separation available, and hence closest to the inertial range. In the conventional logarithmic plot (figure 12(a)), the data approaches $t^3$ scaling from above (see also §3.2). In the range $t/\tau_\eta > 30$, a fit yields $\langle r^2 \rangle \sim t^{2.6}$, which, furthermore, is consistent with our model at the same Reynolds number (figure 11(a)). This implies $\langle r^2 \rangle^{1/3} \sim t^{0.87}$.

In figure 12(b) the same data is shown in a linear plot of $\langle r^2 \rangle^{1/3}$ versus time (black line). Furthermore, the blue and red curves show power laws with exponents 1 and 0.9 respectively. The latter is a better fit to the data in the range $30 < t/\tau_\eta < 130$. This temporal range covers almost an order of magnitude, as is suggested for a reliable power law estimate in §3.1. Note that $t/\tau_\eta < 20 = t_B/\tau_\eta$ is associated with the Batchelor scaling range for $r_0 = 24\eta$, so we do not expect a good fit there. The range $70 < t/\tau_\eta < 130$ is too short to make a distinction between the two power laws. Any smooth curve is piecewise linear over a short range, but that does not mean that the true exponent is equal to 1. Importantly, the exponent 0.9 is consistent with the result from the logarithmic plot, i.e. $\langle r^2 \rangle^{1/3} \sim t^{0.87}$.

Therefore, power laws should be fitted over sufficiently range ranges in a logarithmic plot (as discussed in §3.1) or a linear plot. When this condition is satisfied, the results from the different plots are consistent.

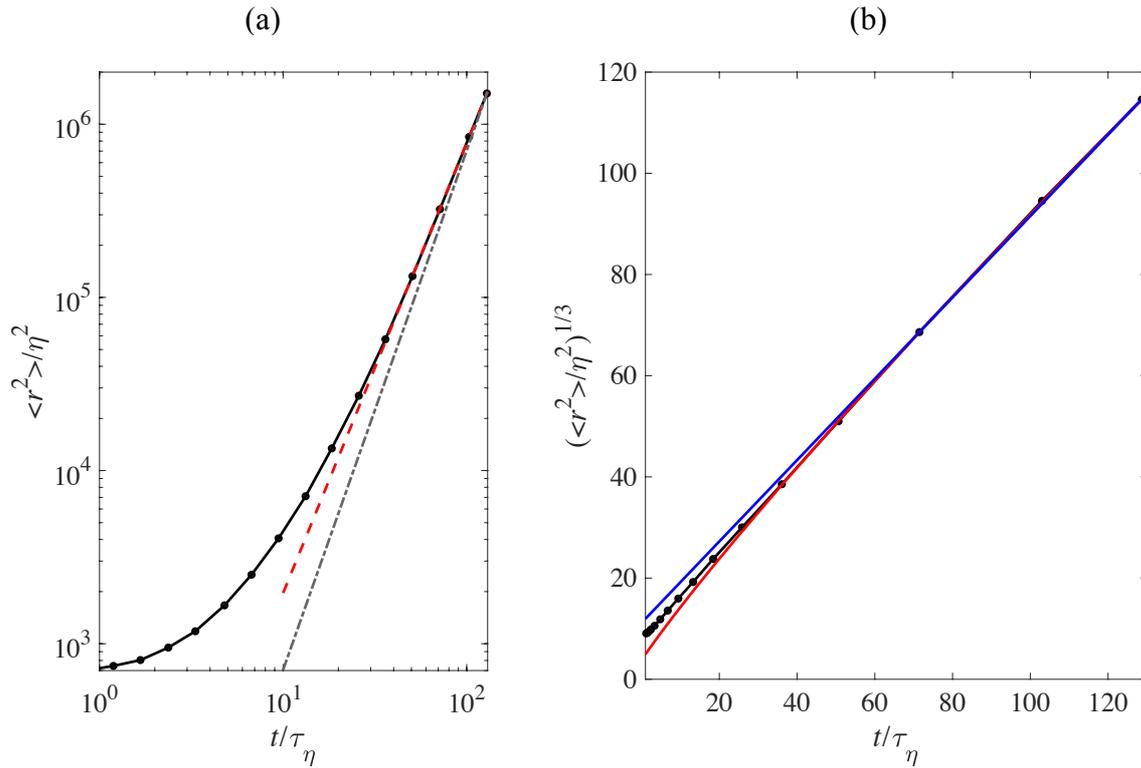

Figure 12. (a) The logarithmic plot of $\langle r^2 \rangle$ versus time for $r_0 = 24\eta$ and $Re_\lambda = 730$ (black line, data source: Bitane, Homann & Bec 2013). The red dashed line and the grey dash-dotted line indicate $t^{2.6}$ and $t^3$ scaling respectively. (b) The linear plot of the same data (black line) compared to power laws with exponents 1 (blue) and 0.9 (red).

# A note on

# "Non-local dispersion and the reassessment of Richardson's $t^3$-scaling law"

G.E. Elsinga[1], T. Ishihara[2] AND J.C.R. Hunt[3]

[1]Laboratory for Aero and Hydrodynamics, Department of Mechanical, Maritime and Materials Engineering, Delft University of Technology, 2628CD Delft, The Netherlands
[2] Graduate School of Environmental and Life Science, Okayama University, Okayama 700-8530, Japan
[3]Trinity College, Cambridge CB2 1TQ, United Kingdom

Correspondence to: g.e.elsinga@tudelft.nl

In Elsinga, Ishihara & Hunt (2022), particle-pair dispersion was discussed based on several datasets including those by Buaria, Sawford & Yeung (2015), who used a binning procedure. However, the effect of binning on the dispersion results remained undiscussed and is now clarified in this brief note. Binning refers to the statistical averaging over particle pairs whose initial separation $r_0$ fall into a specified range, i.e. a bin, as opposed to averaging over a single valued $r_0$. It is a means to improve statistical convergence. Binning is inevitable in experiments, but it is also regularly used in numerical studies, although bin sizes do vary greatly. In our paper, we have referred to the nominal $r_0$ rather than the range of initial separations in case of binning, where the nominal $r_0$ is the (logarithmic) average of the considered range.

Varying bin sizes can introduce minor quantitative differences between datasets. Furthermore, when using binning, the mean square separation velocity associated with the Batchelor regime can be different from the second-order structure function, $S_2$, evaluated at the nominal $r_0$. The latter effect may be noticed, for example, from figure 4 in our paper. This plot shows DNS data from Buaria, Sawford & Yeung (2015) who performed averaging over initial separations within the range $[r_0/2, 2r_0]$ (here $r_0$ indicates the nominal $r_0$). By equating the mean square separation velocity at $t = 0$ to $S_2$, an effective $r_0$ can be defined for each bin, as kindly commented by Dr. D. Buaria. For this dataset, the effective $r_0$ is approximately 1.5 times the nominal $r_0$. On the logarithmic scale, as typically used in scaling studies, the difference between nominal and effective is considered small. More importantly, the results are qualitatively very similar to those obtained using much smaller bin sizes. We do not find important differences in the trends as discussed in the paper. Specifically, and most relevant to our argument, exact $t^3$ scaling is seen for a specific initial separation, which is approximately $r_0 \approx 4\eta$. As noted in our paper, a similar observation was made by Sawford, Yeung & Hackl (2008), who did not use any binning and considered initial separations $r_0/\eta =$ ¼, 1, 4, 16, 64 and 256 for most of their plots. Furthermore, Bitane, Homann & Bec (2012) found that the correction term (to a $t^3$ scaling) was zero only for $r_0 \approx 4\eta$. This seems consistent with their data obtained using small bin sizes, i.e. $\pm 1\eta$ around $r_0$ when $r_0 \leq 16\eta$, and using a fine sampling of initial separations, i.e. $r_0/\eta =$ 2, 3, 4, 6, 8, 12, 16 and larger (Bitane, Homann & Bec 2013). Also, the Reynolds number trend as presented in figure 5 of our paper is robustly seen in other data obtained from smaller bins. The consistency in the results shows that meaningful results can be obtained when using binning. However, one should be aware of some minor quantitative differences between datasets, which results in some uncertainty. For example, at $Re_\lambda \approx 650$, the intermediate-range exponent for a nominal $r_0/\eta = 4$ was estimated to be 2.9 for the data presented by Buaria, Sawford & Yeung (2015) using binning, while it was 3.1 for the data presented by Sawford, Yeung & Hackl (2008) without binning (estimated from their figure 4). The lower exponent in the former may be explained by considering that the effective $r_0/\eta$ was approximately 6 (see above) and that the exponent decreases with increasing $r_0$ at a given Reynolds number. Therefore, approximate signs were used to indicate the condition for $t^3$ scaling, i.e. $r_0 \approx 4\eta$. The uncertainty is of the order of the Kolmogorov length scale, $\eta$, which is inferred from the finest available resolution in $r_0$ and the mentioned differences between the nominal and the effective $r_0$.

**Acknowledgement**

The authors acknowledge the useful comments on the aspect of binning by Dr. D. Buaria.